%% file: Paper.tex
\documentclass[12pt,letterpaper]{article}

\usepackage[margin=1in]{geometry}

\RequirePackage[OT1]{fontenc}
\usepackage{natbib}
\usepackage{setspace}
\RequirePackage[colorlinks,citecolor=blue,urlcolor=blue]{hyperref}
\usepackage[english]{babel}
\usepackage{amsthm}
\usepackage{amsmath,bigints}
\usepackage{amsfonts}
\usepackage{amssymb}
\usepackage{graphicx,subcaption}
\usepackage{enumerate}
\usepackage{color}
\usepackage{array}
\usepackage{psfrag,epsf}
\usepackage{url} 
\usepackage[export]{adjustbox}
\usepackage{bm,float}
\usepackage{titling}

\input{macros}


\addto\captionsenglish{%
}
\begin{document}

\thispagestyle{empty}
\baselineskip=28pt
\vskip 5mm
\begin{center} {\Large{\textbf{Estimating water levels in the High Plains Aquifer by synthesizing satellite data with groundwater well observations}}}
\end{center}

\baselineskip=12pt
\vskip 5mm

\begin{center}
\large
Anis Pakrashi$^{1}$, Murali Haran$^{2}$, Shan Zuidema$^{3}$
\end{center}

\footnotetext[1]{
\baselineskip=10pt 
Department of Statistics, The Pennsylvania State University, University Park, Pennsylvania 16802, USA. (email: avp6357@psu.edu)}
\footnotetext[2]{
\baselineskip=10pt 
Department of Statistics, The Pennsylvania State University, University Park, Pennsylvania 16802, USA. (email: murali.haran@psu.edu)}
\footnotetext[3]{
\baselineskip=10pt 
Earth Systems Research Center, Institute for the Study of Earth, Oceans, and Space, The University of New Hampshire, Durham, New Hampshire 03824, USA. (email: shan.zuidema@unh.edu)}

\baselineskip=17pt
\vskip 4mm
\centerline{September 3, 2026}
\vskip 6mm

\begin{center}
{\large{\bf Abstract}}
\end{center}
The High Plains Aquifer (HPA) is a critical water resource in the Central United States, yet its depletion remains a major concern. While the Gravity Recovery and Climate Experiment (GRACE) satellite mission provides large-scale estimates of liquid water equivalent thickness (LWET), its coarse spatial resolution ($\sim24$ km) limits local inference. In contrast, well observations from the National Ground-Water Monitoring Network (NGWMN) offer valuable but spatially sparse measurements of depth-to-groundwater. In this paper, we develop a downscaling framework for the satellite data that integrates the two sources and covariates using a Bayesian hierarchical framework. Our model uses a latent Gaussian Markov Random Field (GMRF)  that describes the  groundwater storage at a high spatial resolution. To address computational issues, we use a basis representation approach specified via the Moran basis. We find that fine-scale covariates like irrigation intensity and precipitation help refine spatial predictions. We thus provide, to our knowledge, the first statistically-rigorous approach for downscaling groundwater information based on GRACE satellite data and NGWMN groundwater measurements. Our approach yields high-resolution estimates of groundwater variations across the HPA from 2002--2022. The resulting fine-scale inference provides valuable insights into groundwater dynamics, highlighting the effects of land use and local extraction patterns.

\baselineskip=16pt

\vspace{1.5cm}


\newpage
\doublespacing

\section{Introduction}
\label{sec:intro}
Practical social problems often require analyzing data at finer-resolution spatio-temporal regions. The process of transitioning data from a higher (or finer) to a lower (or coarser) resolution is known as aggregation \citep{roquette2018relevance,paige2022spatial}. However, statistical models based on aggregated data introduces certain natural disadvantages \citep{pollet2015taking}. In various research domains such as forestry, geology, agronomy, meteorology, public health, epidemiology, and soil science, information aggregation poses a significant challenge \citep{van1999aggregation,rudstrom2002data}. Finer-resolution information is often obscured due to aggregating data, making intricate trends invisible. The limitations associated with models for aggregated data drive the need for methodologies to recover the finer information from coarser resolution observations. This reverse process is referred to as spatial downscaling or disaggregation, with examples of applications provided by \cite{mertens1997spatial} and \cite{muhling2018potential}. Spatial disaggregation finds applications in hydrology \citep{ALBER2011343}, census data \citep{ijgi8080327}, environmental science  \citep{segond2007simulation}, agriculture \citep{you2009generating}, disease study \citep{arambepola2022simulation}, health \citep{utazi2019spatial}, and other fields.

The High Plains Aquifer, also known as the Ogallala Aquifer, is a vital water resource underlying parts of eight states in the central United States \citep{Qi2010HighPlains}. This natural water source supports nearly one-third of the nation's groundwater irrigation. Geographically, it is subdivided into the Northern, Central, and Southern High Plains (NHP, CHP, and SHP, respectively). The aquifer was formed between approximately 24 and 2 million years ago, during the Miocene and early Pliocene epochs, through the accumulation of debris eroded from the uplifting Rocky Mountains. This material, initially deposited as extensive alluvial fans and later supplemented by windblown sediments, resulted in spatial heterogeneity in hydraulic conductivity. Historical recognition of the aquifer's water depletion (\cite{Haacker2016-to}, \cite{RHODES202383}) predates modern concerns about water security, and informed management of the resource requires a comprehensive understanding of its geological and hydrological evolution.

Satellite gravimetry has become a crucial tool for observing water level changes across both spatial and temporal scales. Since its launch in 2002 as a joint mission between NASA and the German Aerospace Center (DLR), the Gravity Recovery and Climate Experiment (GRACE) has significantly advanced the study of mass transport within the Earth system. The mission enabled continuous monitoring of key processes such as terrestrial water storage, ice sheet and glacier dynamics, sea level fluctuations, and ocean bottom pressure changes \citep{Tapley2019-im}. Originally conceived as a geodetic experiment, GRACE has evolved into a source of consistent and reliable data products, supporting comprehensive assessments of redistribution of surface and subsurface mass variations. The continuation of this effort through the GRACE Follow-On (GRACE-FO) mission extended this record of Earth’s mass variability, allowing the study of long-term trends in the global water cycle. However, while these satellite data are invaluable for tracking regional terrestrial water storage, they are inherently limited by a coarse spatial resolution that makes it challenging to capture localized variations across space and time. Such limitations hinder detailed hydrological analyses, particularly in regions with significant spatial heterogeneity. For instance, while there are emerging indications of improved water conservation \citep{deines2019quantifying} and stabilized storage \citep{STEWARD201636} in recent years across the broader domain, the available coarse-resolution data lacks the fine-scale granularity necessary to characterize localized anomalies. Thus, there remains a critical need for higher-resolution insights to determine how these broader hydrological shifts manifest at a sub-regional level \citep{mehrnegar2023making}.

Another important information source for the in situ groundwater measurements are the network of groundwater wells. Reliable, accurate, and prolonged groundwater monitoring has become increasingly important as groundwater resources are under new pressures from increased energy demand, continued use by agriculture, a focus on stream and wetland ecosystem health, and the impacts of environmental change. Groundwater monitoring at local, regional, and national scales is a key element of responsible  groundwater resource management. As conceived in 2009, the National Ground-Water Monitoring Network (NGWMN) \citep{Wireman2015-ey} was envisioned as a voluntary, cooperative, integrated system of data collection, management, and reporting. It consists of selected groundwater monitoring sites included in established groundwater monitoring networks. Although an important source of local information on water stoarge, the data from this source is too discrete and sparse to provide information on a broad spatial scale. 

Bayesian hierarchical models have become indispensable for analyzing complex stochastic processes, particularly when joint distributions are analytically intractable or high-dimensional. By decomposing intricate systems into nested conditional structures with levels for priors and hyperpriors, they provide a robust framework for uncertainty quantification \citep{schmid2000bayesian}. These models account for multiple sources of variability, yielding consistent estimates across diverse domains, including the medical sciences \citep{yang2022classification,li2023combined}, atmospheric and geosciences \citep{berliner2000long, wainwright2016hierarchical}, ecology \citep{wikle2003hierarchical,ponciano2009hierarchical}, and demography \citep{Pakrashi2025}. Bayesian latent variable modeling \citep{Bishop1998} is particularly advantageous as it assumes that disparate data sources are conditionally independent given the latent process. Furthermore, treating these latent variables as the primary targets of inference \citep{Latent2012} allows for a direct assessment of the underlying physical or social phenomena, thereby providing more nuanced answers to the guiding scientific questions.

A prominent subclass is the Bayesian Latent Gaussian Model, which assigns Gaussian prior densities to latent parameters to facilitate inference in high-dimensional settings \citep{Hazra2023, Hrafnkelsson2023}. However, in large-scale spatial applications, these often have high computational cost associated with the inversion of dense Gaussian covariance matrices. To mitigate this bottleneck, Gaussian Markov Random Fields (GMRFs) \citep{kunsch1979gaussian} offer a computationally efficient alternative by exploiting the sparsity of the precision matrix. Computational tractability can be further enhanced through spatial basis function representations for dimension reduction \citep{cressie2022basis}. This approach significantly accelerates inference while maintaining the structural integrity of the spatial field.

Our study focuses on enhancing the spatial resolution of GRACE water storage data over the HPA by integrating in-situ groundwater measurements with satellite information. The integration of multi-source data remains a fundamental challenge in modern statistics, necessitated by the rapid expansion of data availability from diverse observational platforms and the increasing demand for high-fidelity inference in complex real-world applications (\cite{hector2024efficient}, \cite{Yang02072020}, \cite{LohrRaghunathan}). To account for regional heterogeneity, we incorporate geospatial covariates, including irrigation intensity and precipitation measures. By employing a hierarchical Bayesian framework, this integrated approach generates downscaled estimates of aquifer water storage, enabling a high-resolution assessment of multi-year groundwater dynamics. The modeling framework utilizes latent variable structures, Gaussian Markov Random Field (GMRF) priors, and basis function representations. The ultimate goal is to facilitate the rigorous integration of areal and point-source data, transforming disparate observations into a spatially continuous and geophysically meaningful dataset. 

The remainder of this chapter is organized as follows. Section~\ref{sec:HPA} introduces the High Plains Aquifer, establishing the geographic and hydrological context for the study. Section~\ref{sec:Data} details the satellite product and localized groundwater observations, alongside the covariates and the primary objectives of this multi-source approach. The statistical methodology designed to manage data complexity and reduce computational bottlenecks is developed in Section~\ref{sec:model}, while Section~\ref{sec:final} synthesizes these tools into a final hierarchical Bayesian framework. We explore a novel perspective on second-order moments to characterize volatility within the time series in Section~\ref{sec:volatility}, followed by a presentation of our empirical findings in Section~\ref{sec:result}. Finally, Section~\ref{sec:conclusion} provides concluding remarks and summarizes the broader implications of this work.

\section{Objective}
\label{sec:Objective}

The native resolution of GRACE satellite data ($\sim24$\,km $\times$ 24\,km) is insufficient to resolve localized, fine-scale variations in groundwater storage. Conversely, while well observations provide high-accuracy point data, their spatial sparsity precludes the generation of continuous spatial fields. Our objective is to provide a downscaling framework for the coarse satellite product over the High Plains Aquifer (HPA) that integrates satellite signals with sparse point observations and high-resolution covariates. Our objective is to construct a unified spatial model that synthesizes these multi-source datasets into a continuous, finer-scale spatial field at a resolution of $10$\,km $\times$ 10\,km for each temporal window. Thus, our aim is to yield a panel of shifting temporal windows, each with a high resolution spatial map. 

We employ a Bayesian hierarchical framework to model the fine-resolution latent process, utilizing the computational efficiency of Gaussian Markov Random Fields. To further resolve the computational burden associated with high-dimensional spatial data, we utilize a spatial basis function representation, which enables Bayesian inference within a reduced-dimensional subspace. In addition to estimating the primary groundwater storage depletion, we extend this analysis to model the root mean square (RMS) volatility across each temporal window. This secondary analysis characterizes regions of high localized instability, providing critical insights into areas experiencing significant fluctuations in groundwater storage levels over time, as opposed to the ones with inherent stability.

\section{The High Plains Aquifer}
\label{sec:HPA}

\begin{figure}[t!]
  \centering
    \includegraphics[width=1\textwidth]{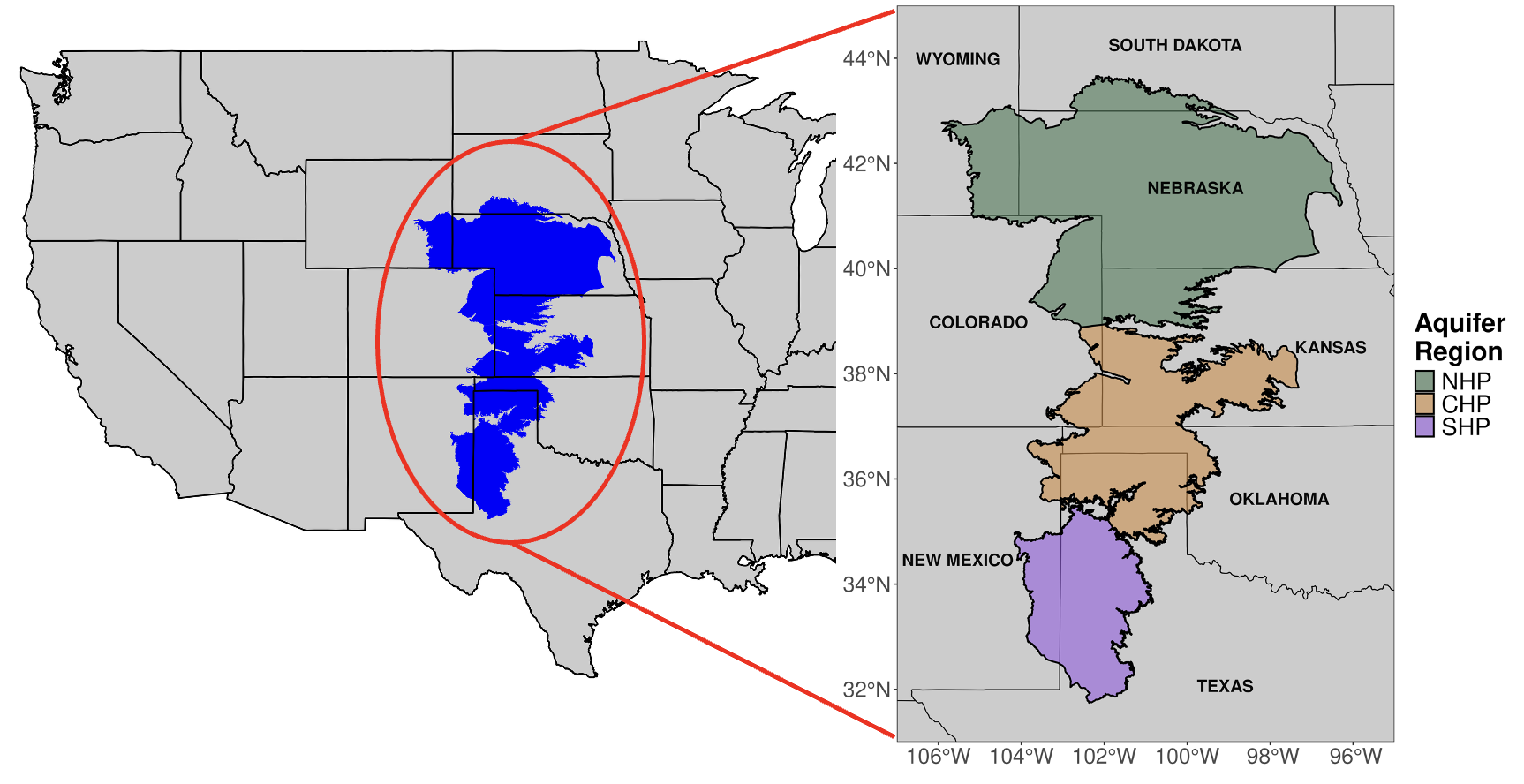}
  \caption{Location of the HPA in the continental United States. The corresponding states are marked along with the aquifer regions -- Northern High Plains (NHP), Central High Plains (CHP) and Southern High Plains (SHP).}
  \label{fig:HPAonUSMap}
\end{figure}

The High Plains aquifer constitutes a critical hydrogeological system spanning approximately 174,000 square miles across the US Midwest \citep{Qi2010HighPlains}. It serves as the primary water source for one of the country's most productive agricultural corridors. Geologically, the aquifer is comprised of multiple hydraulically interconnected units originating from the late Tertiary or Quaternary periods, which are situated atop a bedrock complex ranging in age from the Permian to the Tertiary. For analytical and management purposes, the aquifer is divided into the northern, central, and southern High Plains regions. The northern High Plains aquifer underlies parts of Colorado, Kansas, Nebraska, South Dakota, and Wyoming. The central High Plains underlies parts of Colorado, Kansas, New Mexico, Oklahoma, and Texas. The southern High Plains aquifer underlies eastern New Mexico and northwestern Texas (see Figure \ref{fig:HPAonUSMap}).

In recent decades, the High Plains aquifer has undergone substantial depletion, driven largely by unviable groundwater extraction for large-scale irrigation. This trend is particularly acute in the central and southern regions, posing a direct threat to the long-term agricultural stability of this multi-state area \citep{STEWARD201636}.

\section{Sources of Data}
\label{sec:Data}

We utilize two data sources -- satellite (Section \ref{subsec:GRACE}) and observational sites (Section \ref{subsec:GW}).

\subsection{GRACE Satellite Data}
\label{subsec:GRACE}

The Gravity Recovery and Climate Experiment (GRACE) is a joint venture between NASA and the German Aerospace Center (DLR) that utilizes twin satellites launched in March 2002 to quantify Earth’s mass redistribution by monitoring temporal variations in the global gravity field \citep{Wahr1998-jn}. This initiative, along with its 2018 Follow-On mission (GRACE-FO), facilitates high-resolution tracking of terrestrial water storage, cryospheric mass balance, and sea-level fluctuations resulting from freshwater influx \citep{Landerer2020}. The temporal variations in Earth's gravity field observed on a monthly basis are fundamentally driven by the dynamic redistribution of mass within the Earth system. This mass flux is conceptually modeled as being concentrated within a superficial layer near the planetary surface, typically extending to a depth of several kilometers, and is quantitatively expressed as a change in ``equivalent water thickness'' (EWT). In practice, these gravitational fluctuations primarily originate from terrestrial water storage variations within hydrologic reservoirs, as well as the continuous mass exchange between the atmosphere, oceans, and cryosphere. Although the horizontal scales of these mass shifts extend over hundreds to thousands of kilometers, their vertical magnitude is relatively minute, typically measured in centimeters of EWT. 

In this context, these observed large-scale variations in equivalent water thickness provide a direct window into groundwater recharge, the fundamental hydrologic process by which surface water infiltrates downward through the soil profile to replenish underlying aquifer reserves. Because GRACE integrates the net effect of mass shifts within the subsurface, it serves as a powerful tool for monitoring this replenishment. By linking satellite-derived mass changes to these targeted infiltration frameworks, we can better quantify efforts to restore depleted water levels, mitigate land subsidence, and secure water availability during drought conditions.

In this study, we utilize the CSR GRACE/GRACE-FO RL06.3 Mascon Solutions (RL0603) \citep{GraceMascons} data. To align the spatial resolution of the mascon cells with our specific domain, we focus on all mascons that intersect the High Plains Aquifer (HPA). While our downstream modeling accounts for the complete spatial extent of these intersecting cells, we visually clip the data in our figures to match the exact irregular boundaries of the aquifer. Consequently, our plots display water storage variations constrained strictly within the HPA boundaries. This dataset provides monthly terrestrial water storage (TWS) anomalies expressed in terms of EWT. The satellite outputs are provided at 24 km$^2$ resolution grid, allowing for a coarse spatio-temporal structure of groundwater dynamics. The EWT values are represented as deviations from a baseline. Here, we define baseline as the long-term time average of values in that location. To evaluate the temporal shift in water storage, we analyzed a 21-year period spanning from 2002 to 2022. Our methodology employs a sliding window approach, utilizing 11-year temporal increments to capture the gradual evolution of storage trends across the aquifer. Within each temporal window, we calculate the slope of a linear regression for every spatial pixel.

\begin{figure}[t!]
  \centering
    \includegraphics[width=1\textwidth]{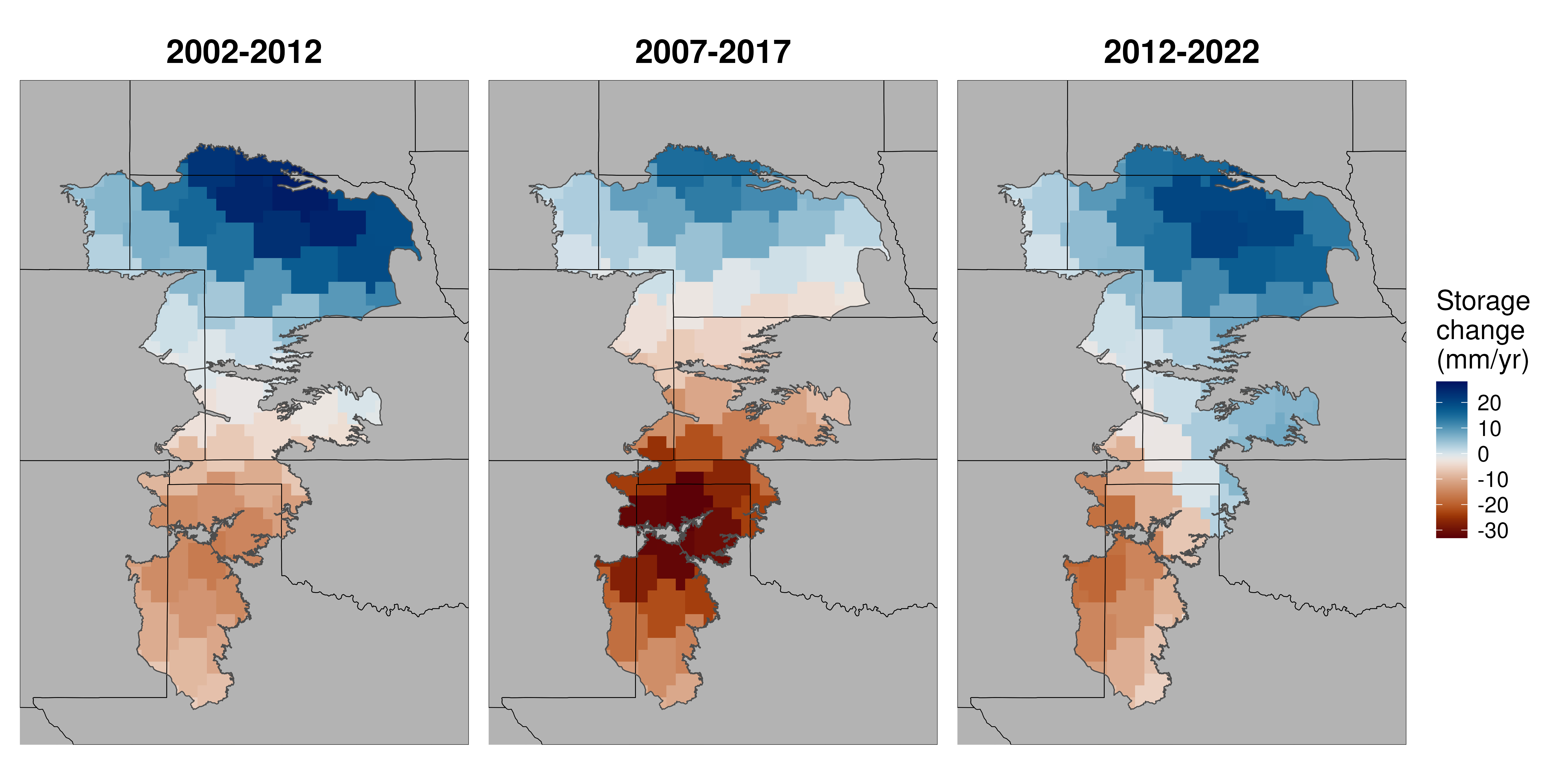}
    \caption{Spatio-temporal variations in water storage across the High Plains Aquifer, visualized through 11-year sliding windows (2002--2022). For simplicity, three temporal windows are displayed here. This represents the coarse resolution information obtained from GRACE satellite data.}
  \label{fig:Grace}
\end{figure}

For a given temporal window at each spatial pixel, let $(y_1, t_1), \ldots, (y_m, t_m)$ represent the $m$ pairs of observations and corresponding time points. We first characterize the linear trend by fitting a regression line:
\begin{equation}\label{eq: slope_GR}
    y_i = \hat{\beta}_0 + \hat{\beta}_1 t_i, \quad i = 1, \dots, m
\end{equation}
where the estimated slope $\hat{\beta}_1$ serves as our primary quantity of interest for long-term storage changes. This pixel-wise trend analysis provides a robust indicator of the localized rate of change in water storage across the HPA over the past two decades. Figure \ref{fig:Grace} provides a condensed visualization of the aforementioned trends.  By considering 11-year windows we assume that short-term variation in other component stores of TWS (e.g. soil moisture, lakes, wetlands) remains static, and the remaining trends can be predominately assocated with groundwater storage change.  However, persistent trends in these other water stores could introduce irreducible errors in our analysis that we neglect.  The plots reveal the Central and Southern High Plains have experienced substantial storage depletion, while the Northern region has accumulated groundwater over the past several decades. 

\subsection{Groundwater Well Observations}
\label{subsec:GW}


The National Ground-Water Monitoring Network (NGWMN) serves as a centralized repository integrating groundwater datasets from local, state, tribal, and federal agencies \citep{WunschSchreiber2024}. By employing a consensus-based framework for data standardization, the network ensures national consistency in reporting groundwater availability and quality. The NGWMN prioritizes long-term observations within principal aquifers, providing the empirical foundation required for multi-scale hydrological modeling and the strategic management of groundwater resources.


\begin{figure}[t!]
  \centering
    \includegraphics[width=1\textwidth]{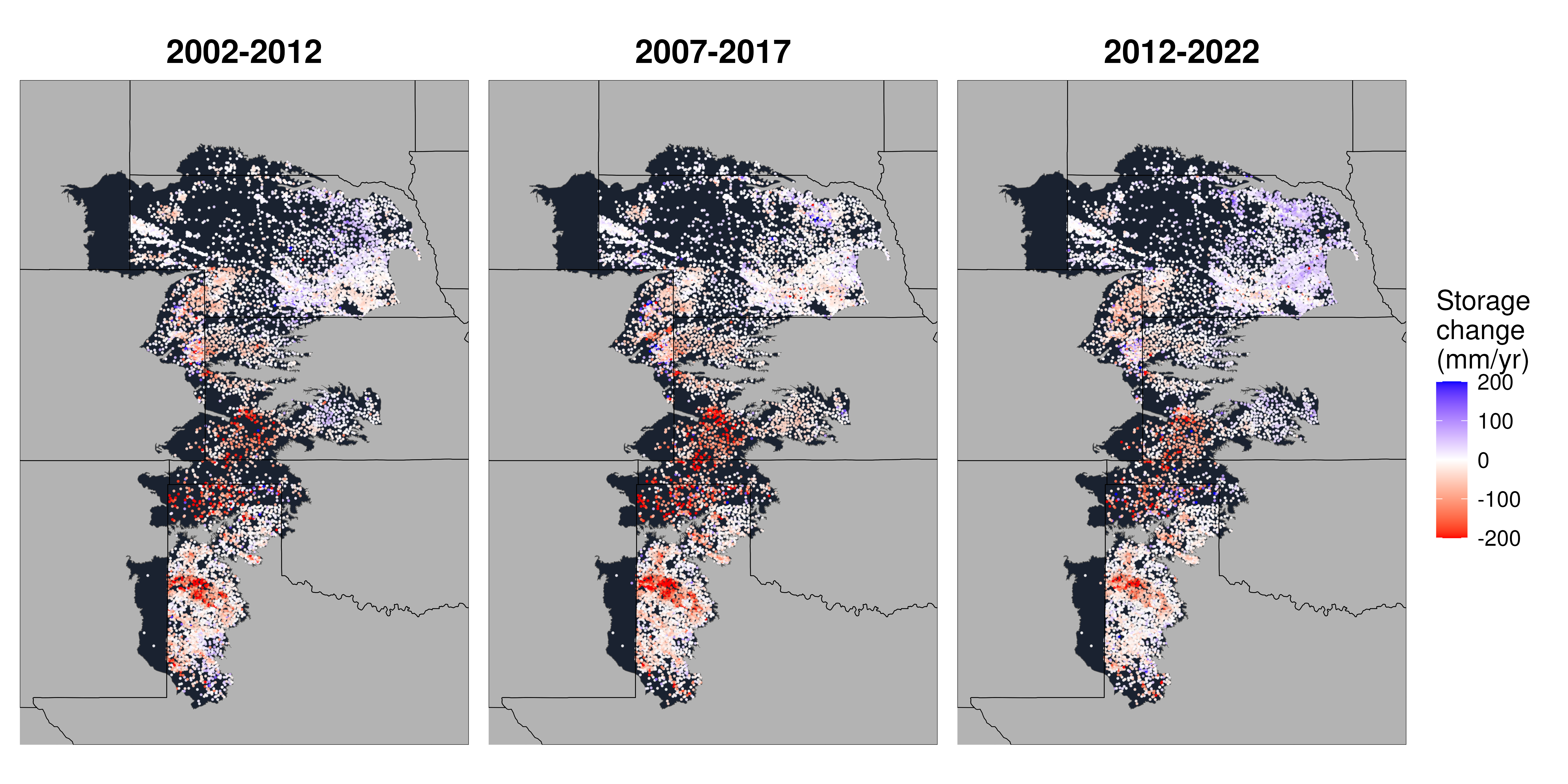}
    \caption{NGWMN groundwater storage changes across the HPA, visualized through 11-year sliding windows (2002--2022). For simplicity, three temporal windows are displayed here. Each point on the map is a well site. For visualization clarity, observations are winsorized to the range $[-200, 200]$ mm/year and the background has been colored in dark.}
  \label{fig:GW}
\end{figure}

In the context of the High Plains Aquifer (HPA), our analysis utilizes 9869 wells integrated into the NGWMN. These in-situ observations provide accurate measurements of depth-to-groundwater levels at specific coordinate points. First, these groundwater observations are multiplied by the specific yield \citep{Liu2022SpecYield} for the corresponding location (See Appendix \ref{append:SY}) to calculate the actual volume of water that changing level reprsesents as either added to or removed from the aquifer. We follow a similar strategy as in Section \ref{subsec:GRACE}, by analyzing the same 21-year time period using a 11-year temporal sliding window approach. As before, we calculate the slope of a linear regression for every site, within each temporal window. As a result, following the procedure described in Equation~\ref{eq: slope_GR}, let us have $m$ pairs of observations and corresponding time points for a given temporal window at each groundwater observation site. Again, we have the linear trend by fitting a regression line with the estimated slope serving as our primary quantity of interest for long-term storage changes (see Figure \ref{fig:GW}).

However, the inherently discrete nature of well observations poses significant challenges for regional spatial characterization. Because monitoring wells are irregularly distributed, they lack the spatial continuity required to construct a representative surface of the aquifer's hydrological state. Furthermore, relying entirely on point-source data introduces the risk of the Modifiable Areal Unit Problem (MAUP), where statistical inferences and observed trends can shift substantially depending on how these discrete points are spatially aggregated or grouped \citep{Jelinski1996, MAUPGeography}. To maximize the utility of the well data, we employ an approach that aligns with the general framework outlined by \cite{Haacker2016-to} to derive a continuous surface from discrete site observations. First, potential outliers, defined as measurements exceeding three standard deviations ($3\sigma$) from the mean, were removed for each well. The localized water level changes were then interpolated to the target resolution using ordinary kriging with an exponential semivariogram model \citep{Isaaks1989AppliedGeostatistics}. While this smoothing step successfully prevents localized noise from introducing spurious artifacts into the model, it is important to note that the resulting surface remains inherently bound to localized variations.

\subsection{Covariates}
\label{subsec:covariates}

Integrating covariate information is essential for improving predictive precision and enhancing model performance across statistical analyses, randomized trials, and machine learning contexts \citep{manski2001designing, brynjarsdottir2014covariate}. In spatio-temporal modeling, covariates provide critical localized insights into spatial and temporal shifts, allowing for the evaluation of local-scale dynamics that may be obscured by broader, more general model trends. While our primary datasets consist of sparse groundwater observations and coarse satellite imagery, incorporating spatially smooth auxiliary data regarding water usage and recharge is vital for refining our model’s inferential power.

\begin{figure}[t!]
  \centering
    \includegraphics[width=1\textwidth]{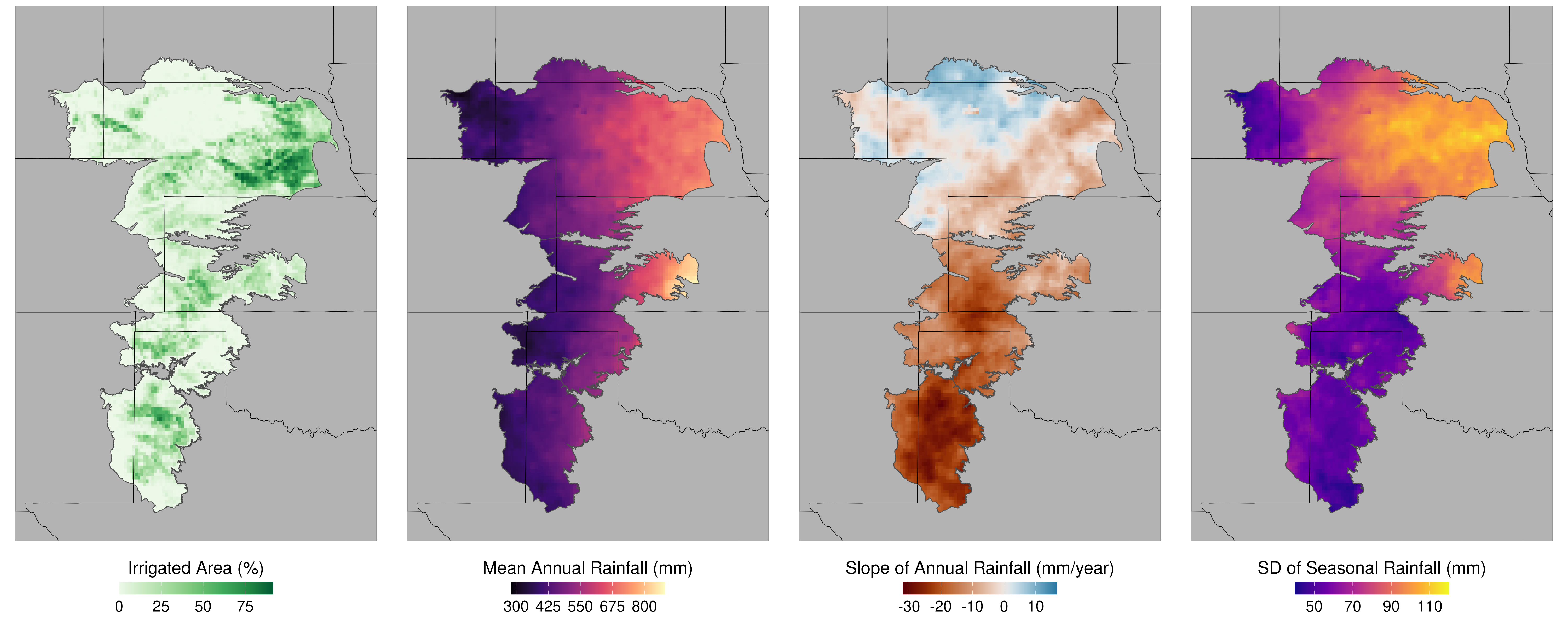}
    \caption{Spatial distribution of covariates for the $2003$--$2013$ study period, including irrigation intensity and precipitation-derived measures.}
  \label{fig:Covariates}
\end{figure}

To this end, we incorporate four covariates into our analysis. The first is irrigation intensity, which represents the percentage of irrigated land within each pixel block \citep{pervez2010irrigation,shrestha2019moderate}. Irrigation represents a primary driver of groundwater drawdown, but also incidental recharge, and consequently this covariate serves as a critical proxy for anthropogenic usage patterns within the High Plains Aquifer region. The remaining three covariates are derived from the gridMET dataset, which provides high-resolution daily surface meteorological data for the contiguous United States \citep{gridMET}. These measures are included because precipitation is the primary driver of groundwater recharge, tracking the natural replenishment of subsurface aquifers. The derived precipitation measures include: (i) mean annual rainfall (mm), which provides an annual picture of precipitation across the HPA; (ii) the slope of annual rainfall (mm/year), representing the annual trend in water availability; and (iii) the seasonal standard deviation of rainfall (mm), which quantifies variability in precipitation patterns across seasons which defines the ability for a portion of precipitation to become recharge during wetter seasons. Figure \ref{fig:Covariates} represents the covariate distribution for the temporal window of $2003$--$2013$.

\section{Model Specifications and Computational Efficiency}
\label{sec:model}

In this section, we explain the statistical components and theoretical frameworks that combine to form the Bayesian hierarchical model. By formalizing these statistical properties, we establish the mathematical foundation necessary to integrate multiple data sources. Section \ref{subsec:bayesian} discusses the theory behind latent variable modeling, Section \ref{subsec:GMRF} explains the concepts related to Gaussian Markov random fields, and Section \ref{subsec:basis} introduces the principles of basis functions, all three of which contribute to efficient computing.

\subsection{Bayesian Latent Variable Hierarchical Model}
\label{subsec:bayesian}

Latent variable models \citep{everett2013introduction} offer a flexible framework for characterizing complex spatial dependencies across a wide variety of data structures. We adopt a hierarchical spatial modeling (HSM) approach \citep{wikle1998hierarchical}, which decomposes the problem into three distinct layers: data, process, and parameters. The data model layer specifies the conditional probability distribution of the observed measurements given the latent spatial Gaussian random fields and associated data-level parameters. The process model layer defines the conditional distributions for the Gaussian spatial processes, governed by process-level parameters or underlying physical dynamics. Finally, the parameter model layer accounts for the uncertainty in the hyperparameters through their respective prior distributions. This hierarchical structure allows for a robust quantification of uncertainty while facilitating the integration of multi-source datasets.

Let $\{Y(\mathbf{s}) : \mathbf{s} \in \mathcal{D}\}$ represent a set of observations indexed by their spatial locations $\mathbf{s}$ within a continuous domain $\mathcal{D} \subset \mathbb{R}^2$. These measurements are obtained across a finite collection of $N$ sites, denoted as $\mathbf{Y} = \{Y(\mathbf{s}_1), \dots, Y(\mathbf{s}_N)\}$. We assume $\mathbf{Y}$ follows a probability distribution $F(\cdot)$, where the choice of $F(\cdot)$ is dictated by the data characteristics—for instance, a Poisson distribution for discrete counts, or Gamma and Beta distributions for bounded or positive continuous variables. The foundation of a Hierarchical Spatial Model (HSM) is a spatially-explicit linear predictor $\eta(\mathbf{s}_i) = \mathbf{X}(\mathbf{s}_i)^\top \boldsymbol{\beta} + \omega(\mathbf{s}_i) + \epsilon(\mathbf{s}_i)$. Here, $\eta$ represents a transformation of the mean of $F$, $\mathbf{X}(\mathbf{s}_i)$ denotes a $p$-dimensional vector of covariates at location $\mathbf{s}_i$, and $\boldsymbol{\beta}$ is the corresponding vector of coefficients. The term $\epsilon(\mathbf{s}_i)$ accounts for independent and identically distributed (IID) error, typically modeled as $\epsilon(\mathbf{s}_i) \sim \mathcal{N}(0, \tau^2)$, where the variance $\tau^2$ represents the ``nugget'' effect or non-spatial micro-scale variation. Spatial correlation is integrated through the random effects $\omega(\mathbf{s}_i)$, which are characterized as a zero-mean Gaussian process. This process is defined by a covariance function $K(\mathbf{s}_u, \mathbf{s}_v)$ with parameters $\boldsymbol{\Theta}$, such that $\boldsymbol{\Sigma}(\boldsymbol{\Theta})_{uv} = K(\mathbf{s}_u, \mathbf{s}_v)$. For any finite set of coordinates $\{\mathbf{s}_1, \dots, \mathbf{s}_n\} \in \mathcal{D}$, the vector of spatial effects $\boldsymbol{\omega} = (\omega_1, \dots, \omega_n)$ follows a multivariate normal distribution $\boldsymbol{\omega} \mid \boldsymbol{\Theta} \sim \mathcal{N}(\mathbf{0}, \boldsymbol{\Sigma}(\boldsymbol{\Theta}))$, where $\boldsymbol{\Sigma}(\boldsymbol{\Theta})$ is an $n \times n$ covariance matrix.

The comprehensive hierarchical structure \citep{haran2024latent} is defined as:
\begin{equation} \label{eq:hsm_framework_1}
\begin{aligned}
    \text{Data Model:} \quad & \mathbf{Y} \mid \cdot \sim F(\Psi, \Theta) ; \quad h(\Psi) := \eta = \mathbf{X} \boldsymbol{\beta} + \boldsymbol{\omega} + \epsilon \\
    \text{Process Model:} \quad & \boldsymbol{\omega} \mid \Theta \sim \mathcal{N}(0, \boldsymbol{\Sigma}(\Theta)) ; \quad \epsilon \sim \mathcal{N}(0, \tau^2 \mathbf{I}) \\
    \text{Parameter Model:} \quad & \boldsymbol{\beta} \sim p(\boldsymbol{\beta}), \quad \Theta \sim p(\Theta), \quad \tau^2 \sim p(\tau^2)
\end{aligned}
\end{equation}
In this framework, $\Psi = h^{-1}(\boldsymbol{\eta})$ relates the mean to the linear predictor via the inverse link function $h^{-1}(\cdot)$. Furthermore, it is standard practice to assign independent prior distributions to the hyperparameter set \citep{banerjee2003hierarchical}.

In this study, we utilize the hierarchical Bayesian framework to integrate multi-source data across distinct modeling layers. Our target spatial domain is discretized at a $10$\,km $\times$ 10\,km resolution, resulting in a latent field of close to $20,000$ pixels. Such high-dimensional settings present significant computational challenges, primarily due to the prohibitive cost of evaluating likelihoods that require inverting and calculating log-determinants of large covariance matrices \citep{rue2005gaussian, williams2006gaussian}. Furthermore, performing Bayesian inference on a latent dimension of this magnitude is traditionally infeasible for standard sampling algorithms. To overcome these bottlenecks, we employ a Gaussian Markov Random Field (GMRF) prior on the latent layer to exploit sparsity in the precision matrix, thereby accelerating likelihood evaluations. To address the dimensionality of the latent space, we utilize spatial basis functions, which project the high-dimensional process onto a lower-dimensional subspace, facilitating efficient Bayesian computation while maintaining high-fidelity spatial inference \citep{wikle2010low, bradley2016comparison}.

\subsection{Gaussian Markov Random Field}
\label{subsec:GMRF}

Directly specifying spatial dependence via $\boldsymbol{\Sigma}(\boldsymbol{\Theta})$, as in Equation~\ref{eq:hsm_framework_1}, generally requires measuring proximity based on distance metrics. In the context of areal data, this approach can be challenging due to the irregular geometries of the regions involved. Additionally, the computational cost of managing dense covariance matrices in high-dimensional settings often becomes a limiting factor. An alternative strategy is a conditional specification, which assumes that the random variable associated with a specific region is primarily influenced by its immediate neighbors. While simple adjacency often defines these neighborhoods, more complex structures may be implemented to suit the specific requirements of the study. Let the spatial random effects $\boldsymbol{\omega}$ in the process model layer be defined conditionally. For a given $s_i \in \mathcal{D}$, let $\boldsymbol{\omega}_{-i}$ denote the vector of spatial effects excluding $\omega(s_i)$. Following \cite{haran2011gaussian}, the full conditional distribution is given by:
\begin{equation} \label{eq:gmrf_dependance}
    \omega(s_{i}) \mid \boldsymbol{\omega}_{-i}, \boldsymbol{\Theta} \sim \mathcal{N} \left( \sum_{j=1}^{n} a_{ij}\omega(s_{j}), \kappa_{i}^{-1} \right), \quad i=1,\dots,n,
\end{equation}
where $\mathbf{A} = ((a_{ij}))$ represents the spatial proximity or adjacency matrix \citep{besag1974spatial, fortin2012spatial}, with $a_{ij} \neq 0$ only if regions $i$ and $j$ are neighbors. Here, $\kappa_i$ represents the precision parameters. To maintain a valid joint distribution, we define the precision matrix $\mathbf{Q}$, based on $\mathbf{A}$, such that $\boldsymbol{\Sigma} = \tau^{-1}\mathbf{Q}^{-1}$. \cite{besag1974spatial} shows that, if $\tau\mathbf{Q}$ is symmetric and positive definite, the process model in Equation~\ref{eq:hsm_framework_1} can be reformulated as:
\begin{equation} \label{eq:gmrf_precision}
    \boldsymbol{\omega} \mid \boldsymbol{\Theta} \sim \mathcal{N}(\mathbf{0}, \tau^{-1}\mathbf{Q}^{-1}(\boldsymbol{\Theta}))
\end{equation}
This formulation leads to the following joint density:
\begin{equation} \label{eq:gmrf_density}
f(\mathbf{\omega} \mid \mathbf{\Theta}) \propto \exp \left( -\frac{1}{2} \mathbf{\omega}^{\top} \mathbf{\Sigma}^{-1} \mathbf{\omega} \right) \propto \exp \left( -\frac{1}{2} \mathbf{\omega}^{\top} (\tau \mathbf{Q}) \mathbf{\omega} \right)
\end{equation}

\noindent
where the inclusion of the precision matrix $\mathbf{Q}$ facilitates direct evaluation of the spatial dependencies by bypassing the need for computationally expensive matrix inversions.

Although the aforementioned GMRF formulation significantly streamlines the likelihood evaluation, a substantial inferential challenge remains. The massive dimensionality of the target resolution presents a parameter space that is virtually impossible to resolve using standard Bayesian inference techniques. This computational bottleneck persists as a primary obstacle, and the model remains computationally intractable until the high-dimensional nature of the latent spatial field is explicitly addressed.

\subsection{Spatial Basis Functions}
\label{subsec:basis}

From a mathematical perspective, a basis for a function space is a collection of elements whose linear combinations can represent any function within that space. As noted by \cite{cressie2022basis}, a countable basis $\{\phi_j(\cdot): j = 1,2,\dots\}$ for the space of square-integrable functions, $\mathcal{F}$, possesses the property that any function $f(\cdot) \in \mathcal{F}$ may be expressed as:
\begin{equation} \label{eq:math_basis}
    f(\cdot) = \sum_{j=1}^{\infty} a_j \phi_j(\cdot)
\end{equation}
In a statistical context, let $Y(\cdot)$ represent a stochastic process. By employing a finite rank $r$ representation, the process is modeled as:
\begin{equation} \label{eq:stat_basis}
    Y(\cdot) = \sum_{j=1}^r a_j \phi_j(\cdot) + \delta(\cdot)
\end{equation}
where $\delta(\cdot)$ is a stochastic process independent of the basis-function coefficients $\{a_j\}$. This term accounts for the residual error or fine-scale variation incurred by the finite truncation of the basis expansion.

In an effort to achieve a sparse reparameterization for areal spatial models, \cite{griffith2003spatial} proposed augmenting the generalized linear framework with select eigenvectors derived from the operator $(\mathbf{I} - \mathbf{11}'/n)\mathbf{A}(\mathbf{I} - \mathbf{11}'/n)$. Here, $\mathbf{I}$ denotes the $n \times n$ identity matrix and $\mathbf{1}$ represents an $n$-dimensional vector of ones. This specific operator is central to the numerator of Moran’s $I$-statistic:

\begin{equation}
    I(\mathbf{A}) = \dfrac{n}{\mathbf{1}'\mathbf{A1}} \dfrac{\mathbf{Y}'(\mathbf{I} - \mathbf{11}'/n)\mathbf{A}(\mathbf{I} - \mathbf{11}'/n)\mathbf{Y}}{\mathbf{Y}'(\mathbf{I} - \mathbf{11}'/n)\mathbf{Y}},
\end{equation}
which serves as a widely utilized non-parametric metric for spatial autocorrelation \citep{moran1950notes}. 

Restricted spatial regression provides a parsimonious alternative by augmenting the fixed effects $\mathbf{X}\boldsymbol{\beta}$ with an efficient basis expansion. Specifically, \cite{hughes2013dimension} introduced the Sparse Areal Mixed Model (SAMM), utilizing the linear predictor $\mathbf{X}\boldsymbol{\beta} + \boldsymbol{\Phi\gamma}$. In this formulation, $\boldsymbol{\Phi}$ is an $n \times q$ matrix composed of the $q$ principal eigenvectors of the Moran basis, and $\boldsymbol{\gamma} \sim \mathcal{N}\{\mathbf{0},(\tau\boldsymbol{\Phi}'\mathbf{Q}\boldsymbol{\Phi})^{-1}\}$ represents the spatial random effects. The matrix $\mathbf{Q} = \text{diag}(\mathbf{A1}) - \mathbf{A}$ corresponds to the Laplacian of the underlying connectivity graph \citep{brouwer2011spectra}. 

Spatial filtering typically employs the eigenvectors of the Moran operator, $M_1 = (\mathbf{I} - \mathbf{11}'/n)\mathbf{A}(\mathbf{I} - \mathbf{11}'/n)$. As argued by \cite{hughes2017spatial}, significant dimensionality reduction is attainable by restricting the basis to $q \ll n$ vectors. The choice of $q$ is based on the ``best DIC'' criterion. We fit a sequence of models corresponding to a sequence of values for $q$ and use DIC to choose the ‘right’ model, having the least DIC value. Letting these $q$ vectors comprise the columns of $\boldsymbol{\Phi}$, the filtering linear predictor is expressed as:
\begin{equation}\label{eq:classicalspatialfiltering}
    g(\boldsymbol\mu) = \mathbf{X}\boldsymbol{\beta} + \boldsymbol{\Phi\gamma}
\end{equation}
Building upon this, \cite{hughes2017spatial} developed a Bayesian Spatial Filtering (BSF) approach. This method adopts the transformed conditional mean from Equation~\ref{eq:classicalspatialfiltering} and assigns a prior to the spatial coefficients $\boldsymbol{\gamma}$ as follows:

\begin{equation}\label{eq:BSFprior}
    \boldsymbol{\gamma} \mid \tau \sim \mathcal{N}\{\mathbf{0},(\tau\boldsymbol{\Phi}'\mathbf{Q}\boldsymbol{\Phi})^{-1}\}
\end{equation}

In our model, we use basis representation via Bayesian spatial filtering due to high-dimensionality and cross-correlation of spatial random effects. Spatial covariance is approximated using $q$ basis functions, with $\boldsymbol{\omega} \approx \boldsymbol{\Phi}\boldsymbol{\gamma}$, where $\boldsymbol{\Phi}$ is $n \times q$ basis matrix and $\boldsymbol{\gamma}$ the coefficient vector, $q<<n$. Equation~\ref{eq:hsm_framework_1} is modified as follows:
\begin{equation} \label{eq:hsm_framework_sbf}
\begin{aligned}
    \text{Data Model:} \quad & \mathbf{Y} \mid \cdot \sim F(\Psi, \Theta) ; \quad \eta = \mathbf{X} \boldsymbol{\beta} + \boldsymbol{\Phi}\boldsymbol{\gamma} + \epsilon \\
    \text{Process Model:} \quad & \boldsymbol{\gamma} \mid \boldsymbol{\xi} \sim \mathcal{N}(0, \Sigma(\boldsymbol{\xi})) ; \quad \epsilon \sim \mathcal{N}(0, \tau^2 \mathbf{I}) \\
    \text{Parameter Model:} \quad & \boldsymbol{\beta} \sim p(\boldsymbol{\beta}), \quad \boldsymbol{\xi} \sim p(\boldsymbol{\xi}), \quad \tau^2 \sim p(\tau^2)
\end{aligned}
\end{equation}

\begin{figure}[t!]
  \centering
    \includegraphics[width=1\textwidth]{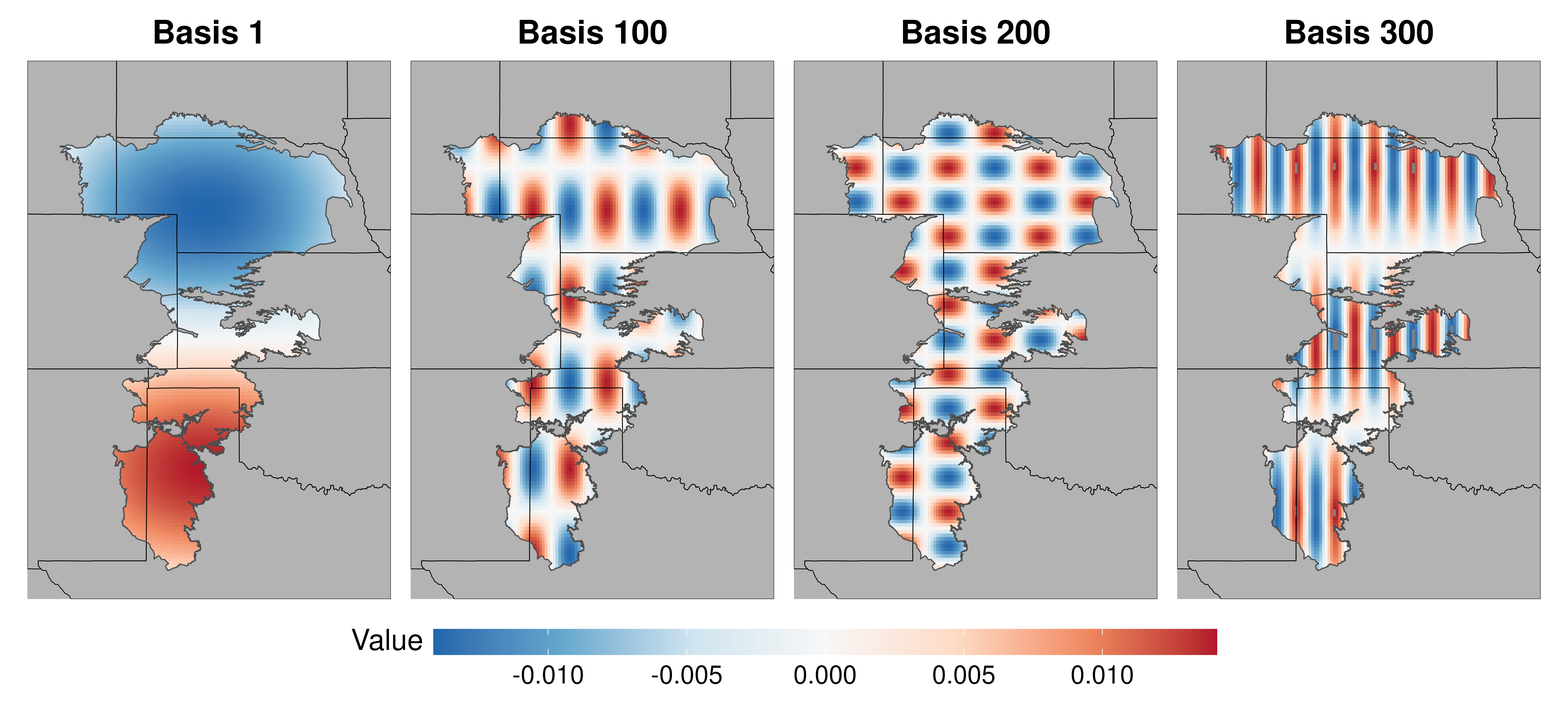}
    \caption{Sample basis vectors for the $2003-2013$ study period.}
  \label{fig:BSF}
\end{figure}

In our application, we set $q = 300$ (See Appendix \ref{append:Basis}). This selection facilitates a massive reduction in dimensionality within the process model, projecting the latent field from close to $20,000$ spatial pixels down to a subspace defined by just $300$ basis vectors. Consequently, the inferential framework is significantly streamlined, requiring the estimation of only $300$ basis coefficients rather than the original high-dimensional latent process. Additionally, the prior specification for the latent layer remains governed by the structure defined in Equation~\ref{eq:BSFprior}. Figure \ref{fig:BSF} shows the $1^{\text{st}}$, $100^{\text{th}}$, $200^{\text{th}}$, and $300^{\text{th}}$ basis vectors for the temporal window of $2003-2013$. We note contrasting patterns, which transition from broad spatial gradients to increasingly finer oscillations as the basis index increases. Together, these multi-scale components effectively capture the localized spatial heterogeneity across the aquifer.

\section{Final Bayesian Hierarchical Model}
\label{sec:final}

To integrate multi-source spatial data while maintaining computational efficiency, we employ a hierarchical Bayesian framework. We define the underlying spatial process at the target resolution as $\omega(s)$, where $s$ corresponds to a location on a fine-scale lattice of $10\text{ km}^2$. Let $\mathbf{Z} = \mathbf{X}\boldsymbol{\beta} + \boldsymbol{\omega}$ denote the latent linear predictor, a vector with dimensions equal to the latent layer.

The model assimilates two distinct data streams. Let $B_i$, for $i= 1, \dots, n_B$, represent the $i^{th}$ low-resolution block from the GRACE dataset, where $n_B$ is the total number of blocks. Let $Y(B_i)$ denote the GRACE satellite slope data for block $B_i$. Since the target latent representation comprises high-resolution units nested within these larger blocks, we assume that $Y(B_i)$ is approximately the average of the $N_i$ latent values of $\mathbf{Z}$ situated within that block:
\begin{equation}\label{eq:GraceBlock}
    Y(B_i) \approx N_i^{-1} \sum_{j \in B_i} Z(s_j)
\end{equation}

Additionally, let $Y(s_k)$ represent the groundwater slope values extracted from the krigged monitoring well data at location $s_k$, for $k = 1, \dots, n_P$, where $n_P$ denotes the total number of cells in the latent layer at our target spatial resolution. These interpolated values across each temporal window, are assumed to be direct realizations of the underlying latent process $\mathbf{Z}$:
\begin{equation}\label{eq:GWBlock}
    Y(s_k) \approx Z(s_k)
\end{equation}

We define $\sigma_B^2$ and $\sigma_P^2$ as the variance components for the GRACE and groundwater model layers, respectively. The complete data model, conditional on the latent linear predictor $\mathbf{Z}$, is specified as follows:
\begin{equation}\label{eq:DataModel}
    \begin{aligned}
        Y(B_i) \mid Z(s_j), \sigma_B^2 &\sim \mathcal{N} \left(N_i^{-1} \sum_{j \in B_i} Z(s_j), \sigma_B^2\right), \quad i = 1, \dots, n_B \\
        Y(s_k) \mid Z(s_k), \sigma_P^2 &\sim \mathcal{N} \left(Z(s_k), \sigma_P^2\right), \quad k = 1, \dots, n_P
    \end{aligned}
\end{equation}

Assuming conditional independence, the joint posterior distribution of the process parameters and hyperparameters is given by:
\begin{equation}\label{eq:JointPost}
    \pi(\mathbf{Z}, \sigma_B^2, \sigma_P^2 \mid \mathbf{Y}) \propto \prod_{i=1}^{n_B} \mathbb{P}[Y(B_i) \mid Z, \sigma_B^2] \times \prod_{k=1}^{n_P} \mathbb{P}[Y(s_k) \mid Z, \sigma_P^2] \times \pi(\mathbf{Z}) \times \pi(\sigma_B^2) \times \pi(\sigma_P^2)
\end{equation}
We assign inverse-Gamma priors to the variance components $\sigma_B^2$ and $\sigma_P^2$ and a gamma prior on the precision parameter $\tau$. Given the structure $\mathbf{Z} = \mathbf{X}\boldsymbol{\beta} + \boldsymbol{\omega}$, we specify independent normal priors for the covariate coefficients $\boldsymbol{\beta}$. To ensure computational tractability, we utilize the spatial basis representation $\boldsymbol{\omega} \approx \boldsymbol{\Phi}\boldsymbol{\gamma}$ as discussed in previous sections. Consequently, the prior specification for the spatial process is reduced to the basis coefficients $\boldsymbol{\gamma}$, for which we assign a Gaussian Markov Random Field prior as detailed in Section~\ref{subsec:basis}.

For the inferential analysis, we employ Bayesian Hamiltonian Monte Carlo (HMC) sampling implemented in \texttt{Stan}. We aim at $4000$ post burn-in samples for each parameter and obtain around similar number of effective samples. To optimize throughput, independent \texttt{Stan} routines are executed in parallel for each temporal window. Computational tasks were performed on the \texttt{Linux}-based RoarCollab high-performance cluster, maintained by the Institute for Computational and Data Sciences (ICDS) at the Pennsylvania State University. All analyses were conducted using the \texttt{R} statistical software environment, with the total computational suite requiring less than a day to reach completion.

\section{Are slopes enough ?}
\label{sec:volatility}

The primary objective of this research is to derive downscaled information at a $10\text{ km} \times 10\text{ km}$ resolution regarding changes in terrestrial water storage. This has been achieved by augmenting the GRACE dataset with point-source groundwater observations and a suite of spatial covariates. Our methodology evaluates these changes over shifting 11-year temporal windows, utilizing the slope of a linear regression at each spatial unit as the principal estimator for the rate of change in stored water.

Beyond the mean trend, the temporal stability of these changes is of significant interest. We investigate the temporal volatility of these processes by analyzing their second moments. To quantify this variability, we employ the root mean square (RMS) of successive differences as a proxy for temporal volatility. In this context, the RMS serves as an intuitive indicator of stability: elevated values highlight regions undergoing abrupt transitions or significant instability in water storage, whereas lower values show areas with more resilient, steady-state dynamics. By adopting this metric, we can more effectively compare regional stability across each 11-year window, distinguishing between areas of persistent change and those that remain relatively constant over time.

Mathematically, let $(y_1, t_1), \ldots, (y_m, t_m)$ represent the $m$ pairs of observations and corresponding time points, for a given temporal window at each spatial pixel or site. To quantify the volatility of storage changes, we calculate the RMS of the successive differences of the observations. This metric is defined as:

\begin{equation}\label{eq:RMS_volatility}
    \text{RMS} = \sqrt{\frac{1}{m-1} \sum_{i=1}^{m-1} (y_{i+1} - y_i)^2}
\end{equation}

This measure allows us to distinguish between regions of stable progression and those experiencing high-frequency fluctuations in water storage levels. We use the formula in Equation~\ref{eq:RMS_volatility} for both the satellite and the site data. The remaining steps in hierarchical modeling are exactly the same as described in Section \ref{sec:final}.

\section{Results}
\label{sec:result}

Our analysis yields a dual-layered inferential output, with downscaled spatial maps providing critical insights into both long-term temporal trends and the temporal volatility of water storage. Section \ref{subsec:slope} presents our findings related to the primary focus of storage changes, Section \ref{subsec:RMS} details the patterns of temporal volatility, and Section \ref{subsec:Compare} compares our results with similar previous studies in the literature.

\subsection{Change in Water Storage}
\label{subsec:slope}

\begin{figure}[b!]
  \centering
    \includegraphics[width=1\textwidth]{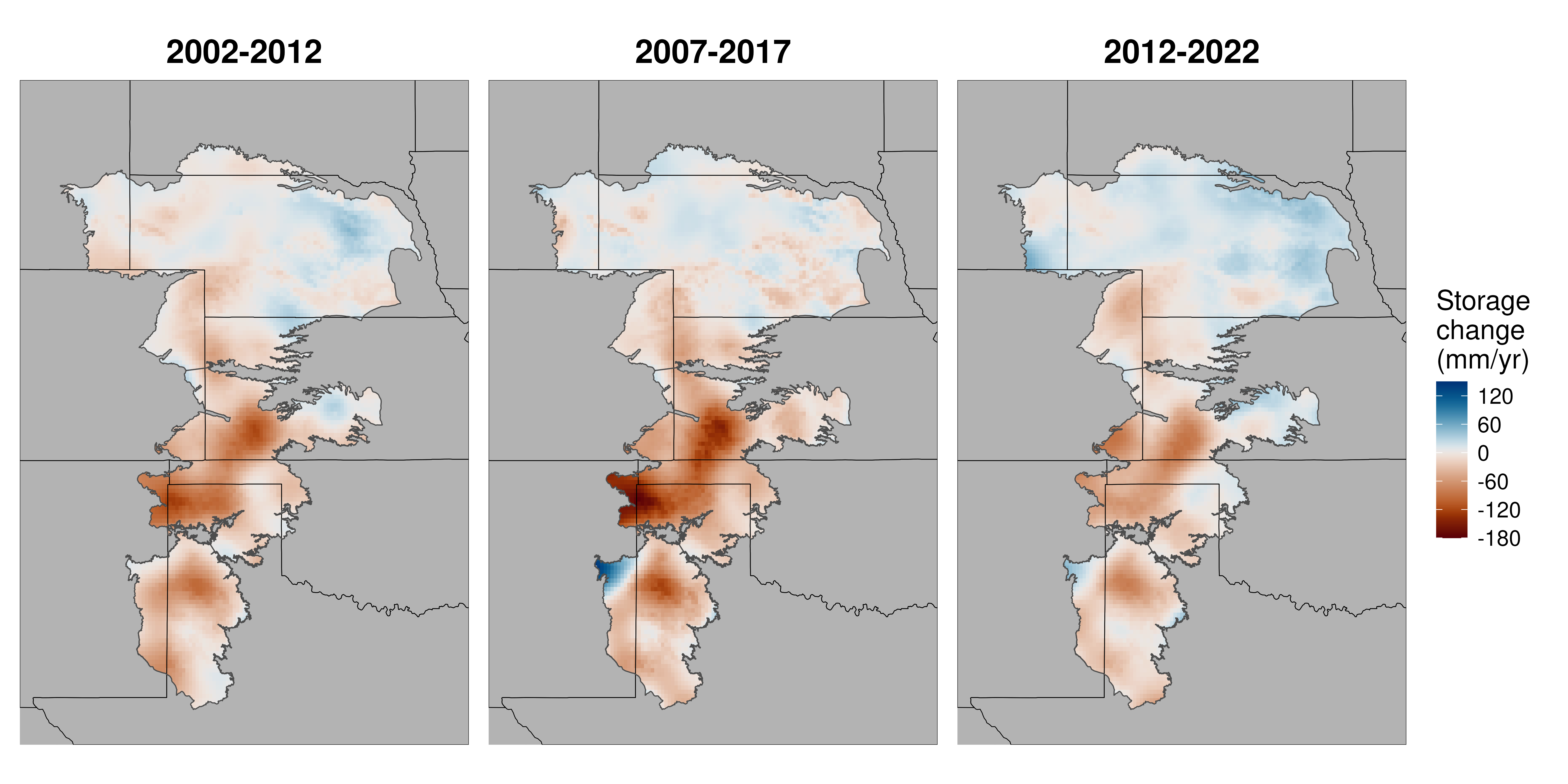}
    \caption{Downscaled spatio-temporal variations in water storage across the High Plains Aquifer, visualized through eleven 11-year sliding windows (2002--2022). For simplicity, three temporal windows are displayed here. This represents the results of our method that combines GRACE information with groundwater site observations and covariates.}
  \label{fig:slope_plot}
\end{figure}

As detailed in Section \ref{sec:final}, we present a detailed final set of maps that shows downscaled features across space. The temporal dynamics are captured through shifting 11-year windows, with the longitudinal results illustrated in Figure \ref{fig:slope_plot}. Our framework reveals fine-scale spatial variations that are otherwise obscured in the coarse-resolution GRACE signal. The addition of valuable local information from groundwater sites and covariates enhanced the downscaled satellite maps.

The results primarily highlight a pervasive decline in water storage, presenting a critical concern for regional water security. Spatially, the most pronounced long-term depletion is concentrated within portions of the Central and Southern High Plains. This substantial drawdown is indicative of intensive groundwater extraction for agricultural irrigation and municipal consumption, a deficit further worsened by the region's semi-arid to arid conditions. Given that the High Plains Aquifer (HPA) serves as the primary hydrological resource in these sectors, an overdependence on stored groundwater has led to a pattern of depletion \citep{sophocleous2005groundwater, steward2013tapping}. In contrast, the extreme Northern High Plains, particularly across parts of Nebraska, exhibited a slight net increase in water storage \citep{scanlon2012groundwater}. This region's proximity to the Missouri River network may suggest that alternative surface water resources help mitigate localized pumping stress on the underlying aquifer. Furthermore, localized recharge via infiltration from adjacent fluvial systems might contribute to the relative stability observed in this northern sector \citep{stanton2010simulation}. Macro-climate factors, such as winter snowfall patterns in Nebraska and the Dakotas spring snowmelt may provide a seasonal source of water that potentially supports natural infiltration before peak summer irrigation begins.

\begin{figure}[b!]
  \centering
    \includegraphics[width=1\textwidth]{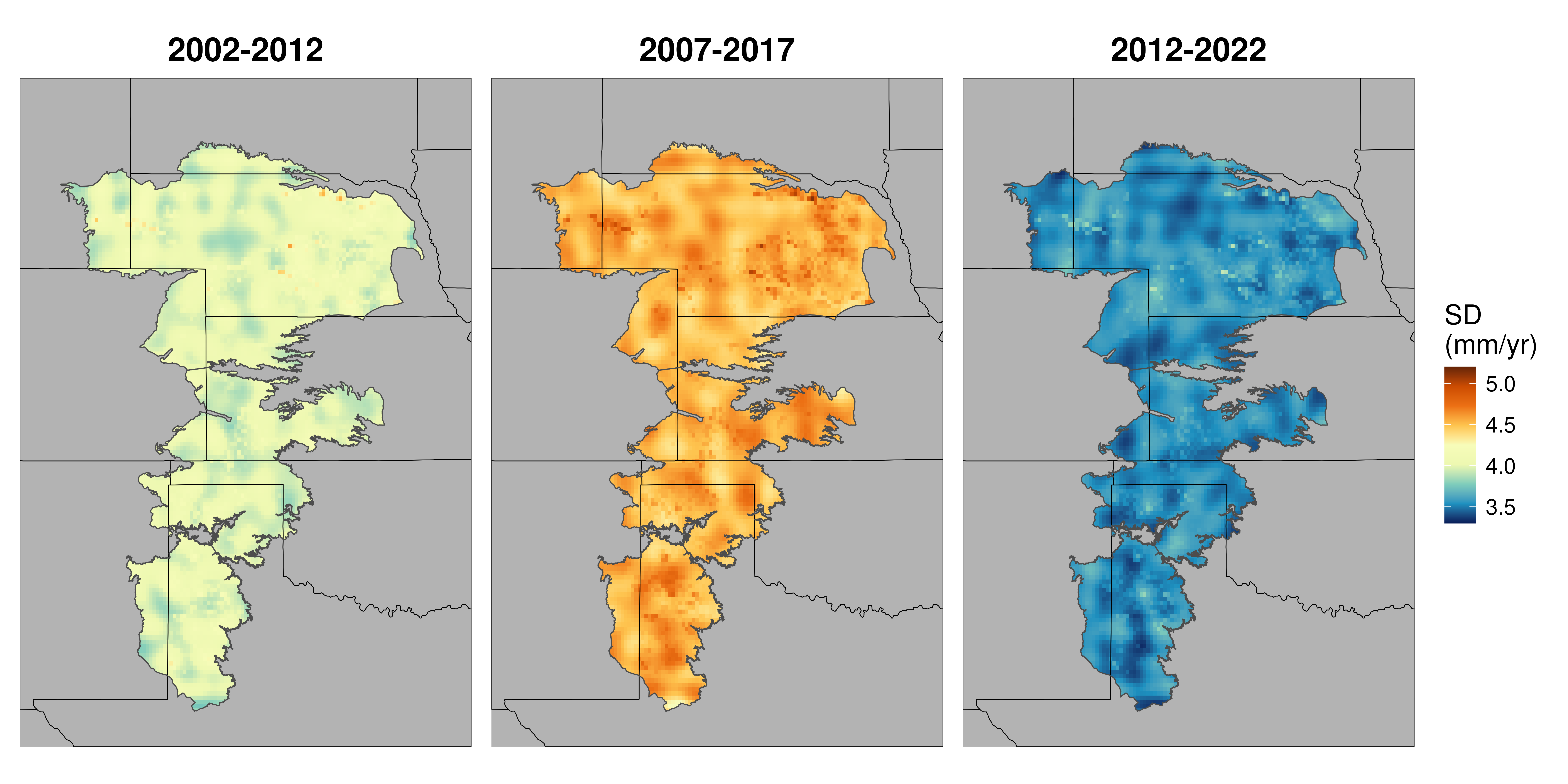}
    \caption{Posterior standard deviation from the analysis reveals regions with different uncertainty levels in obtaining the storage change estimates. Again, for simplicity, three temporal windows are displayed here. }
  \label{fig:sel_postsd_plot}
\end{figure}

While extreme storage declines are evident in the earlier intervals, these trends appear to stabilize in more recent years, potentially reflecting advancements in water-efficient technologies and a heightened awareness of conservation. The most severe depletion across the Central and Southern High Plains was concentrated during the middle of our study period. These acute drops are tied directly to the severe North American drought of 2012--2015, which stands as one of the most significant stressors on the aquifer's storage during these two decades \citep{RIPPEY201557, jacob2012ebsco}. 

Looking at our uncertainty quantification, the posterior standard deviation maps reveal spatial non-uniformity. This variation highlights specific regions where combining information from two distinct data sources naturally yields higher uncertainty in the final estimates. Furthermore, we observe a clear temporal shift in estimation uncertainty that depends on the specific time frame under consideration. This pattern implies that the nature and structural behavior of our estimates evolve across time. These uncertainty maps are presented in Figure \ref{fig:sel_postsd_plot}.


\subsection{Volatility}
\label{subsec:RMS}

\begin{figure}[t!]
  \centering
    \includegraphics[width=1\textwidth]{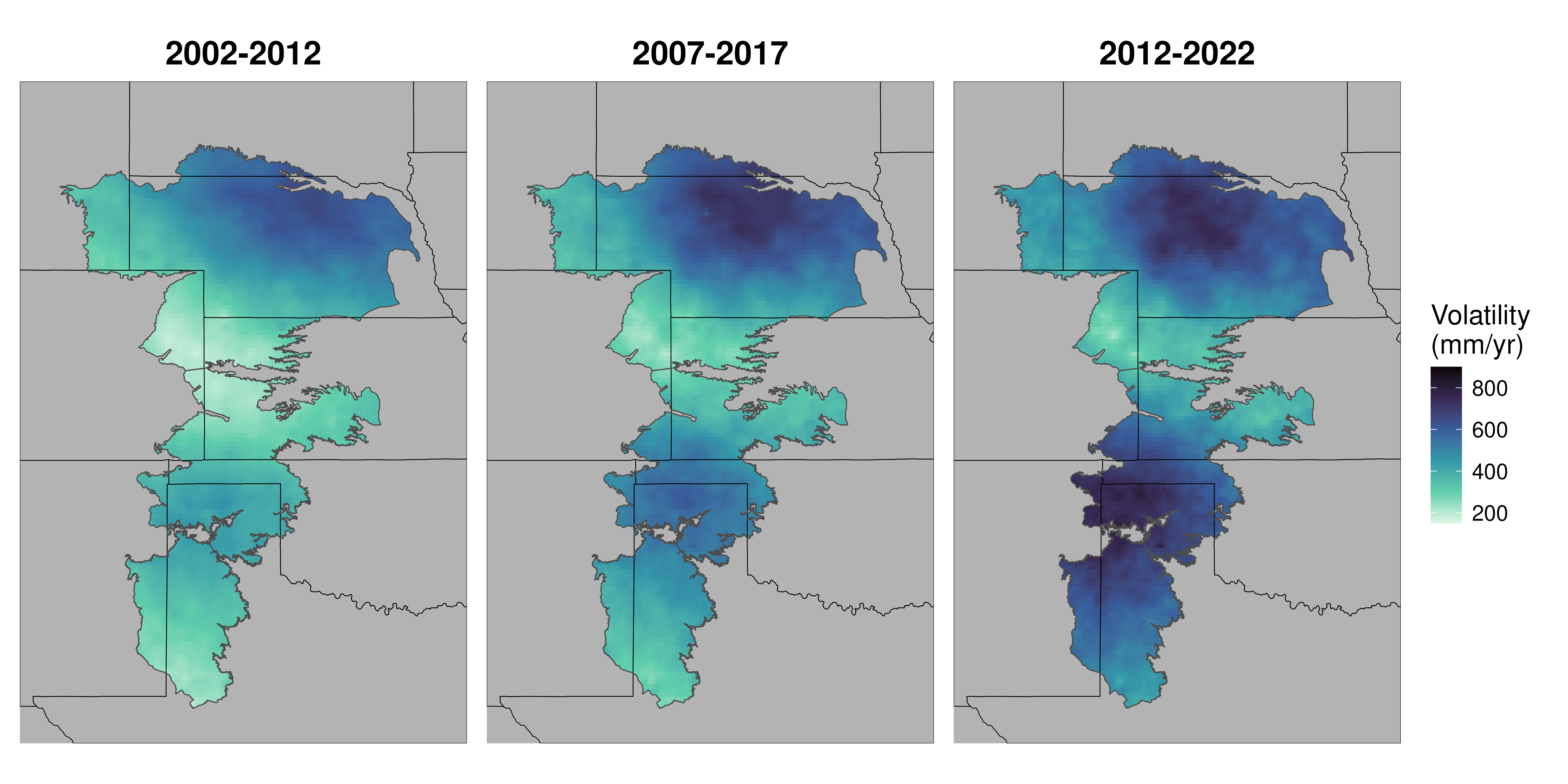}
    \caption{Downscaled temporal volatility in water storage across the High Plains Aquifer, visualized through eleven 11-year sliding windows (2002--2022). For simplicity, three temporal windows are displayed here.}
  \label{fig:vol_plot}
\end{figure}

As detailed in Section \ref{sec:volatility}, the inclusion of second-moment analyses facilitates a comprehensive evaluation of the high-frequency fluctuations prevalent within each temporal window, over and above the long-term storage trends. This measure of volatility reveals that while certain regions maintain relatively stable storage levels, others are characterized by severe transitions throughout the 11-year study periods. Figure \ref{fig:vol_plot} presents the downscaled volatility metrics derived from our Bayesian inferential framework, illustrating that annual fluctuations vary significantly across the High Plains Aquifer.

We observe that certain regions like central Nebraska in the Northern High Plains and the Texas and Oklahoma Panhandles in the Central and Southern High Plains demonstrate the lowest levels of temporal stability. On the other hand, the regions of HPA within the limits of Kansas and Colorado remained relatively stable over time. From a longitudinal perspective, these spatial discrepancies were less pronounced in the earlier stages of the study but escalated notably over time. The most recent decade has experienced the maximum spatial heterogeneity regarding the stability of water storage dynamics. The localized variations in groundwater trends may be attributed to a combination of climatic variability and multi-scale economic drivers that dictate irrigation water demand \citep{scanlon2012groundwater}. Beyond local weather conditions, structural depletion is heavily influenced by agricultural market dynamics, where changes in global population and food consumption shift the demand for commodity crops sourced from major agricultural regions like the US Midwest \citep{haqiqi2023global}. These global trade dependencies and localized economic incentives govern extraction volumes, making groundwater stress an intersection of macro-scale supply chain pressures and physical resource limits \citep{hancock2024texas}.Consequently, our results provide a characterization of water storage change by reporting both aggregate changes and high-resolution spatial features. This approach identifies specific regions experiencing significant fluctuations driven by the local net of recharge and extractions.

\subsection{Comparison with other studies}
\label{subsec:Compare}

\begin{figure}[b!]
  \centering
    \includegraphics[width=1\textwidth]{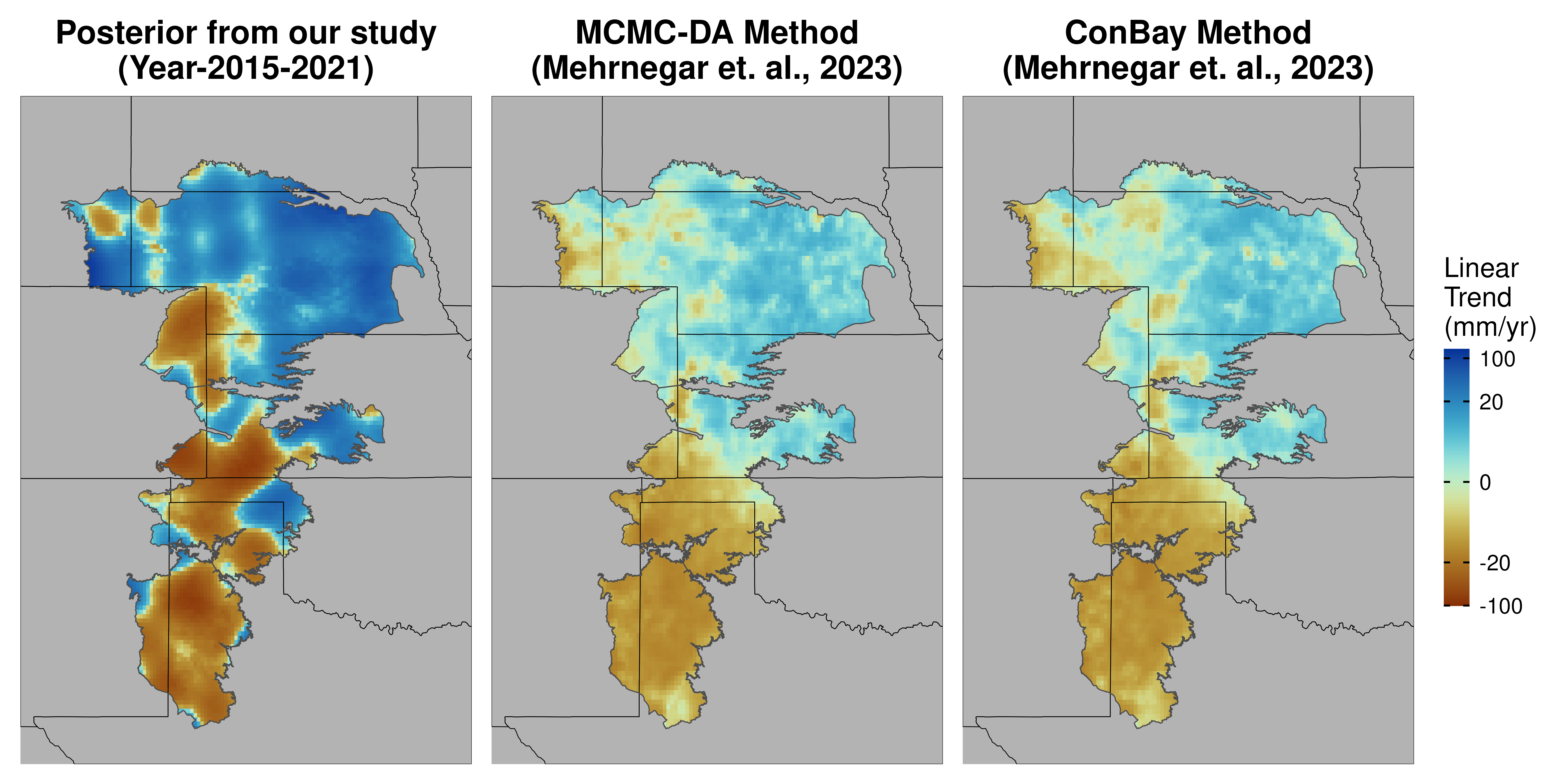}
    \caption{Comparing the spatial features of the high-resolution maps of linear storage change trend among our proposed method and two different methods used in \cite{mehrnegar2023making}, over the temporal window $2015$--$2021$. We use a transformed scale to accommodate the variation in ranges of the different maps.}
  \label{fig:Mehrnegar}
\end{figure}

Our proposed methodology utilizes a novel unified spatial model within a Bayesian hierarchical framework. While a direct methodological equivalent, specifically regarding our temporal integration, is not available in current literature, we evaluate the spatial reliability of our results through comparison with an existing study of similar extent. We compare our findings with the high-resolution terrestrial water storage linear trends ($mm/year$) developed by \cite{mehrnegar2023making}. These trends, illustrated in Figure~\ref{fig:Mehrnegar}, are also derived from GRACE and GRACE-FO satellite data.

Here, we benchmark our estimated linear storage trends (slopes) against the high-resolution terrestrial water storage (TWS) data developed by \cite{mehrnegar2023making} over the 7-year time range ($2015$--$2021$). Specifically, we compare our output to their ``MCMC-DA'' (Markov Chain Monte Carlo with Data Augmentation) and ``ConBay'' (Constrained Bayesian Optimization) implementations. At regional scales, these approaches offer complementary perspectives and yield comparable insights, with all models identifying similar broad geographic regions of storage depletion across the High Plains Aquifer. However, the absolute magnitudes and fine-scale features differ. The data assimilation approach in \cite{mehrnegar2023making} integrates satellite observations with physical land surface models to resolve high-fidelity features driven by soil and hydrologic properties, providing potentially a higher resolution data product. Conversely, our empirical data fusion framework utilizes local monitoring well networks. Consequently, our approach provides greater isolation of localized drawdown patches, particularly across the highly heterogeneous landscapes of the Central and Southern High Plains.

Our results also compare against the groundwater-based dataset from \cite{Haacker2016-to}. Our unified Bayesian spatiotemporal model yields a significantly smoother gradient across the High Plains Aquifer, avoiding the localized patches observed in the \citeauthor{Haacker2016-to} results. These discrepancies are largely attributable to the underlying methodological philosophies. Unlike interpolation-based methods that depend strictly on high physical well density, this empirical framework balances local observations with globally available satellite data from GRACE and GRACE-FO. By utilizing the inherent spatial correlation of the gravity field to merge these information sources, the model mitigates the localized artifacts common in standard interpolations. This approach provides a cohesive representation of storage trends across the study area, though its scalability requires further evaluation across regions with varying monitoring network densities.

\section{Conclusion}
\label{sec:conclusion}

Water resources are fundamental to human survival, the preservation of Earth's ecosystems, and socio-economic stability. Systematic changes in groundwater levels have profound, long-standing impacts on population dynamics, regional security, and the viability of agricultural patterns. Consequently, obtaining precise, high-resolution knowledge of water level trends and stability variations is essential for identifying vulnerable regions that require urgent management.

This study achieves this objective through a rigorous hierarchical Bayesian integration of multi-scale data. We utilized coarse-resolution satellite information from the GRACE mission to identify broad areas of regional variation, integrating it with site-specific groundwater well observations across the High Plains Aquifer (HPA) region to provide exact measures of localized storage changes. To further refine these estimates, we incorporated additional environmental and anthropogenic covariates, such as precipitation patterns and irrigation intensity, allowing the model to capture the complex interplay between groundwater drawdown and natural infiltration.

The results of this study provide critical insights into localized variations in water storage, facilitating evidence-based water management at scales significantly finer than the coarse footprint of original GRACE data. This research demonstrates that the integration of auxiliary in-situ measurements and spatially-informed covariates yields substantial inferential gains, offering a state-of-the-art statistical downscaling framework that directly bridges the gap between global satellite gravimetry and sparse, irregular physical well networks without relying on purely deterministic interpolation. Characterizing these shifts is uniquely urgent for the HPA, where decades of intensive irrigation have caused severe, heterogeneous water table declines. Because the HPA spans vast landscapes with highly localized depletion pockets, coarse data is fundamentally insufficient for effective governance.

Ultimately, this methodology offers a precise diagnostic tool for groundwater management, empowering researchers to investigate and answer critical questions pertaining to storage changes at a granular resolution of just 10 square kilometers. By revealing the actionable spatial heterogeneity of drawdown in the aquifer, these downscaled maps can be utilized as high-fidelity observational datasets for model calibration problems in physical models that operate at a similar spatial scale. By providing a more accurate target for calibration, our findings help reduce data-model bias and enhance the predictive reliability of regional hydrological simulations for long-term resilience.

\section*{Funding Statement}

The work is supported by the US Department of Energy, Office of Science under Cooperative Agreement No. DE-SC0022141. 

\section*{Acknowledgments}

The authors are grateful for the support  of Dr. Rob Nicholas and Dr. Lidiia Iavorivska, director and project manager of the Program on Coupled Human and Earth Systems (PCHES) research group; PCHES  provided the opportunity for this interdisciplinary collaboration. 

\section*{Data Availability Statement}

All data used in this study are publicly available. The two primary datasets can be accessed online: the GRACE Mascons satellite data are available from the Center for Space Research (\url{https://www2.csr.utexas.edu/grace/RL06_mascons.html}), and the groundwater-level data are available from the National Ground-Water Monitoring Network (\url{https://www.usgs.gov/apps/ngwmn/index.jsp}). All processed datasets and associated files have been provided as Data Files.

\section*{Disclosure Statement}

The authors declare no potential conflict of interest.

\baselineskip=14pt


\newpage

\begin{appendix}
\doublespacing

\section{Specific Yield}
\label{append:SY}

\begin{figure}[b!]
  \centering
    \includegraphics[width=0.5\textwidth]{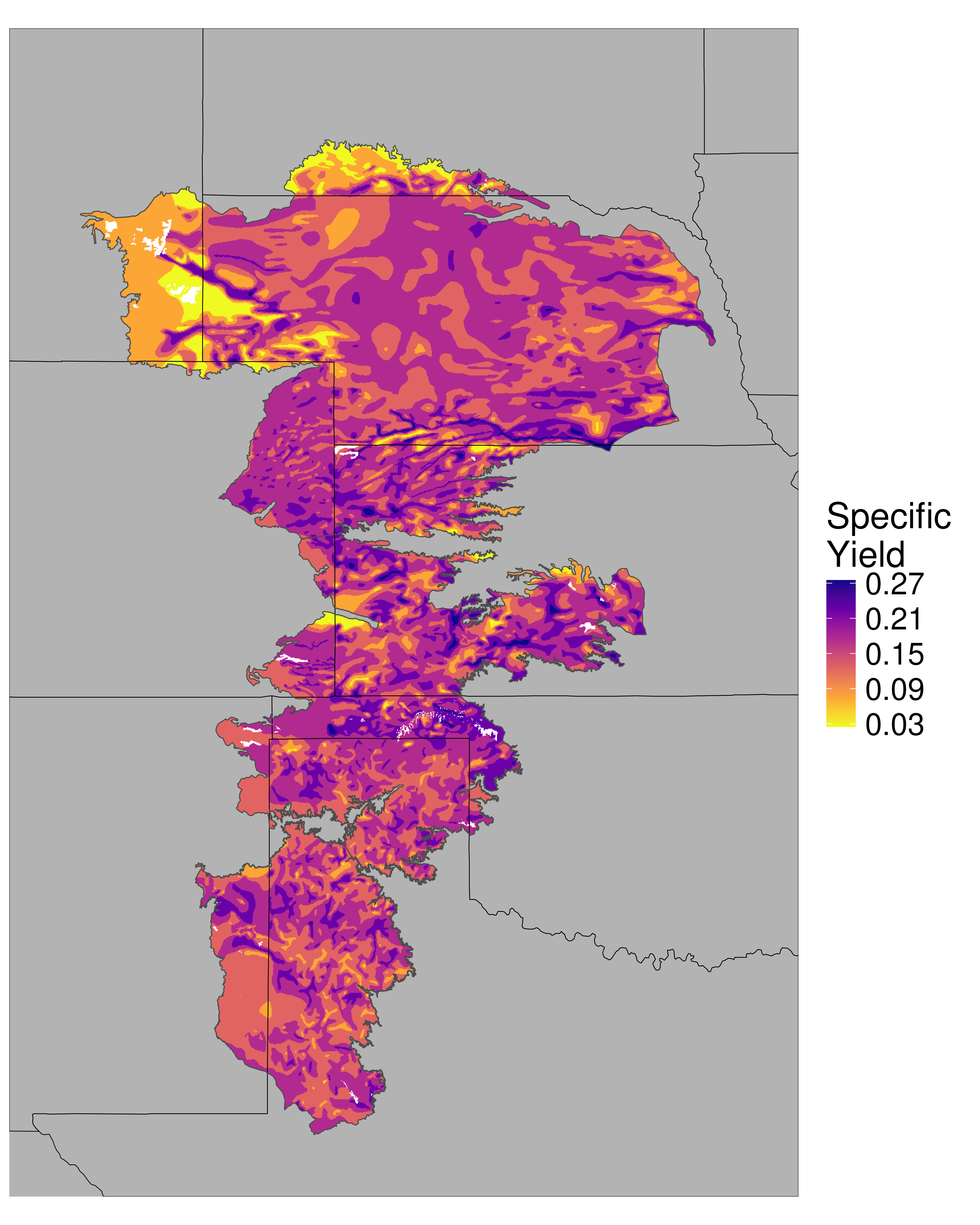}
    \caption{Specific Yield values for the HPA region}
  \label{fig:specificyield}
\end{figure}

To translate raw changes in physical water table depth into an actual equivalent volume of liquid water, hydrogeologists rely on a critical aquifer property known as the specific yield ($S_y$). An aquifer is not an open underground cavern filled entirely with water; rather, it is a complex structural matrix composed of porous soil, sand, and gravel. When the water table drops or rises by a certain vertical distance, the volume of water actually gained or lost from storage does not equal that entire vertical column, as much of the space is occupied by solid sediment, and some residual water remains bound to soil particles by capillary action. Therefore, the specific yield ($S_y$) acts as a dimensionless scaling factor, typically represented as a decimal fraction, that quantifies the actual ratio of drainable water volume to the total volume of the aquifer. Mathematically, the true change in groundwater storage ($\Delta \text{GWS}$) is obtained by multiplying this specific yield by the raw change in the measured groundwater level ($\Delta \text{GWL}$), structured as $\Delta \text{GWS} = S_y \times \Delta \text{GWL}$. This crucial multiplication scales raw structural depth measurements down to a true volumetric mass equivalent, ensuring that physical observations from individual wells reflect actual water storage dynamics (see Figure \ref{fig:specificyield}).

\section{Choice of Number of Basis Vectors}
\label{append:Basis}

To determine the optimal number of spatial basis functions for our analysis, we evaluate model performance across a single temporal window spanning 2003--2013. We construct our basis matrices using the first $q$ eigenvectors of the Moran operator, varying $q$ from 100 to 500 in increments of 50. The resulting Leave-One-Out Information Criterion (LOOIC) values, alongside their corresponding effective number of parameters ($p_{\text{loo}}$) and standard errors, are summarized in Table~\ref{tab:append1}.

\begin{table}[b!]
\centering
\begin{tabular}{rrrrrrr}
    \hline
    q & looic & se\_looic & p\_loo & se\_p\_loo & drop & unit\_drop \\ 
    \hline
    100.00 & 241740.37 & 530.36 & 106.87 & 2.66 &  -  &  - \\ 
    150.00 & 237403.62 & 572.63 & 148.14 & 3.75 & 4336.75 & 173.47 \\ 
    200.00 & 235171.52 & 580.87 & 189.90 & 4.56 & 2232.10 & 89.28 \\ 
    250.00 & 232559.02 & 594.09 & 230.74 & 5.52 & 2612.50 & 104.50 \\ 
    300.00 & 231235.27 & 597.96 & 274.22 & 6.51 & 1323.75 & 52.95 \\ 
    350.00 & 229567.03 & 619.65 & 316.13 & 7.76 & 1668.25 & 66.73 \\ 
    400.00 & 228442.41 & 631.30 & 361.66 & 9.25 & 1124.62 & 44.98 \\ 
    450.00 & 227617.38 & 637.65 & 403.65 & 10.51 & 825.03 & 33.00 \\ 
    500.00 & 226847.11 & 647.57 & 447.39 & 11.90 & 770.28 & 30.81 \\ 
    \hline
\end{tabular}
\caption{Model diagnostic metrics under different choices of basis function dimension $q$, evaluated using the 2003--2013 temporal window.}
\label{tab:append1}
\end{table}

\begin{figure}[t!]
  \centering
  
  \begin{subfigure}[b]{0.5\textwidth}
    \centering
    \includegraphics[width=1\textwidth]{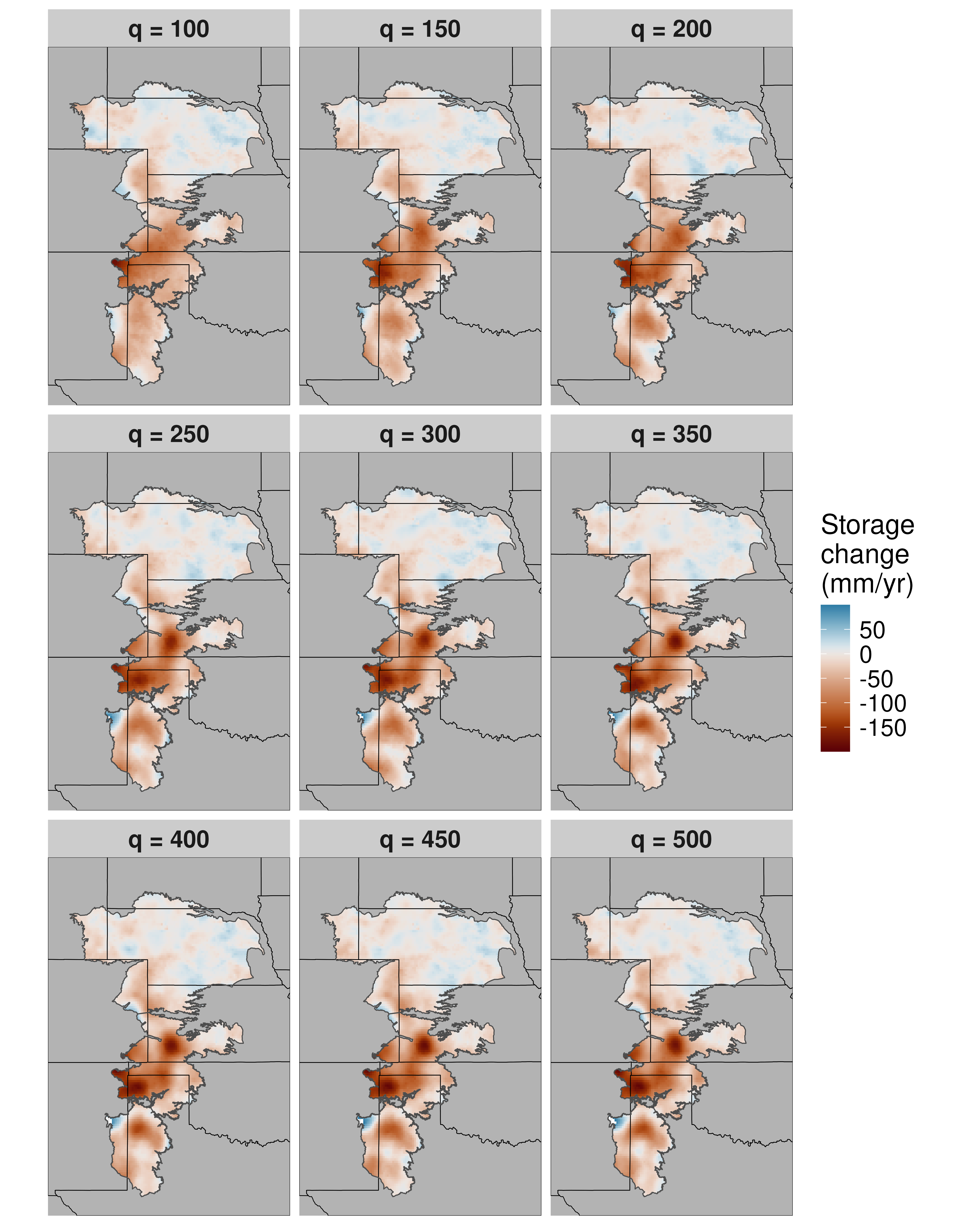}
    \caption{}
    \label{fig:Choice_q}
  \end{subfigure}
  \hspace{0.5cm} 
  \begin{subfigure}[b]{0.45\textwidth}
    \centering
    \includegraphics[width=1\textwidth, height=4.5cm]{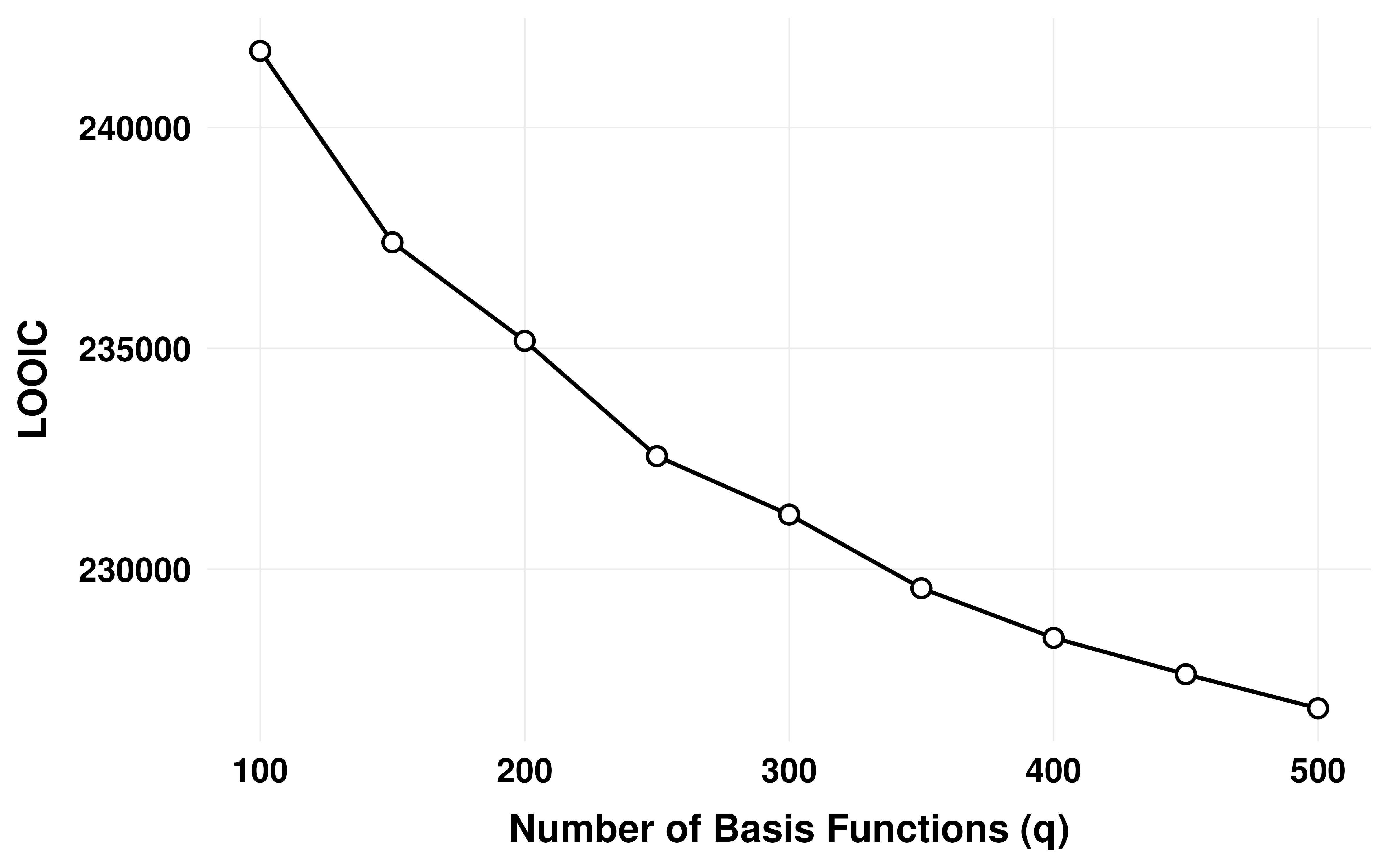}
    \caption{}
    \label{fig:LOOIC_plot}
    
    \vspace{0.3cm} 
    
    \includegraphics[width=1\textwidth, height=4.5cm]{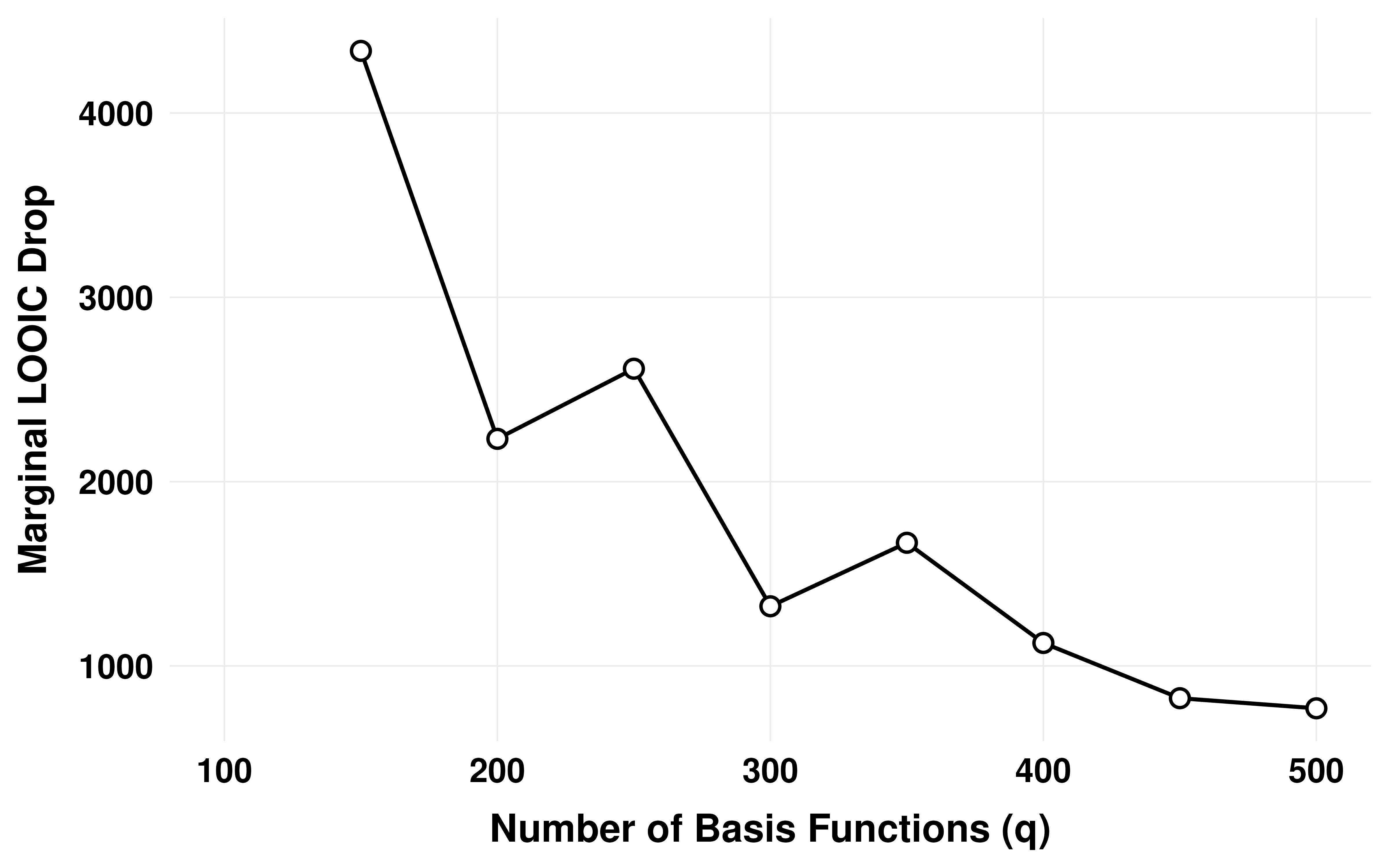}
    \caption{}
    \label{fig:LOOIC_Drop_plot}
  \end{subfigure}

  \caption{(a) Posterior mean estimates of groundwater storage changes ($2003$--$2013$) across a range of basis function dimensions $q$, (b) total LOOIC values plotted against $q$, and (c) marginal decrease in LOOIC between consecutive values of $q$.}
  \label{fig:BasisFnChoice}
\end{figure}

As shown in Figure \ref{fig:LOOIC_plot}, the total LOOIC decreases steadily as the number of basis functions increases, exhibiting a characteristic leveling off that suggests diminishing returns for higher dimensions. To more precisely identify the point of stabilization, we examine the marginal drops between consecutive values of $q$ in Figure \ref{fig:LOOIC_Drop_plot}. These incremental improvements drop sharply initially and begin to stabilize to low values around $q = 300$.

This quantitative choice is further supported by the spatial estimates presented in Figure \ref{fig:Choice_q}, which illustrates the posterior means of the groundwater storage change fields as a function of $q$. At lower values of $q$, critical localized features appear oversmoothed. These spatial structures begin to resolve more clearly around $q = 250$ and remain highly consistent from $q = 300$ onward, with only minor high-frequency variations appearing at higher dimensions. Balancing computational efficiency in our hierarchical model with the need to capture fine-scale spatial features, these combined diagnostics justify our choice of $q = 300$ for the full spatio-temporal analysis.

\end{appendix}

\end{document}

%% file: macros.tex
\def\frac#1#2{{\textstyle{#1\over#2}}}

\DeclareSymbolFont{AMSb}{U}{msb}{m}{n}
\DeclareMathSymbol{\Natural}{\mathbin}{AMSb}{"4E}
\DeclareMathSymbol{\Integer}{\mathbin}{AMSb}{"5A}
\DeclareMathSymbol{\Real}{\mathbin}{AMSb}{"52}
\DeclareMathSymbol{\Rational}{\mathbin}{AMSb}{"51}
\DeclareMathSymbol{\Imaginary}{\mathbin}{AMSb}{"49}
\DeclareMathSymbol{\Complex}{\mathbin}{AMSb}{"43} 
\DeclareMathSymbol{\Disk}{\mathbin}{AMSb}{"44} 
\def\bi{\begin{itemize}}
\def\ei{\end{itemize}}
\def\bd{\begin{description}}
\def\ed{\end{description}}
\def\ben{\begin{enumerate}}
\def\een{\end{enumerate}}
\def\bv{\begin{verbatim}}
\def\ev{\end{verbatim}}

\def\hat#1{{\widehat{#1}}}

\def\2to{{\ {\buildrel 2\over \longrightarrow}\ }}

\def\I1ton{{$I_1,\ldots,I_n$}}
\def\X1ton{{$X_1,\ldots,X_n$}}
\def\Y1ton{{$Y_1,\ldots,Y_n$}}
\def\Z1ton{{$Z_1,\ldots,Z_n$}}
\def\R1ton{{$R_1,\ldots,R_n$}}
\def\e1ton{{$e_1,\ldots,e_n$}}
\def\t1ton{{$t_1,\ldots,t_n$}}
\def\x1ton{{$x_1,\ldots,x_n$}}
\def\y1ton{{$y_1,\ldots,y_n$}}
\def\z1ton{{$z_1,\ldots,z_n$}}

%% file: Paper.bbl
\begin{thebibliography}{}

\bibitem[Abatzoglou, 2013]{gridMET}
Abatzoglou, J.~T. (2013).
\newblock Development of gridded surface meteorological data for ecological applications and modelling.
\newblock {\em International Journal of Climatology}, 33(1):121--131.

\bibitem[Ahmed et~al., 2012]{Latent2012}
Ahmed, A., Aly, M., Gonzalez, J., Narayanamurthy, S., and Smola, A.~J. (2012).
\newblock Scalable inference in latent variable models.
\newblock In {\em Proceedings of the Fifth ACM International Conference on Web Search and Data Mining}, WSDM '12, page 123–132, New York, NY, USA. Association for Computing Machinery.

\bibitem[Alber and Piégay, 2011]{ALBER2011343}
Alber, A. and Piégay, H. (2011).
\newblock {Spatial disaggregation and aggregation procedures for characterizing fluvial features at the network-scale: Application to the Rhone basin (France)}.
\newblock {\em Geomorphology}, 125(3):343--360.

\bibitem[Arambepola et~al., 2022]{arambepola2022simulation}
Arambepola, R., Lucas, T.~C., Nandi, A.~K., Gething, P.~W., and Cameron, E. (2022).
\newblock A simulation study of disaggregation regression for spatial disease mapping.
\newblock {\em Statistics in Medicine}, 41(1):1--16.

\bibitem[Banerjee et~al., 2003]{banerjee2003hierarchical}
Banerjee, S., Carlin, B.~P., and Gelfand, A.~E. (2003).
\newblock {\em Hierarchical modeling and analysis for spatial data}.
\newblock Chapman and Hall/CRC.

\bibitem[Berliner et~al., 2000]{berliner2000long}
Berliner, L.~M., Wikle, C.~K., and Cressie, N. (2000).
\newblock {Long-lead prediction of Pacific SSTs via Bayesian dynamic modeling}.
\newblock {\em Journal of climate}, 13(22):3953--3968.

\bibitem[Besag, 1974]{besag1974spatial}
Besag, J. (1974).
\newblock Spatial interaction and the statistical analysis of lattice systems.
\newblock {\em Journal of the Royal Statistical Society: Series B (Methodological)}, 36(2):192--225.

\bibitem[Bishop, 1998]{Bishop1998}
Bishop, C.~M. (1998).
\newblock {\em Latent Variable Models}, pages 371--403.
\newblock Springer Netherlands, Dordrecht.

\bibitem[Bradley et~al., 2016]{bradley2016comparison}
Bradley, J.~R., Cressie, N., and Shi, T. (2016).
\newblock {A comparison of spatial predictors when datasets could be very large}.
\newblock {\em Statistics Surveys}, 10(none):100 -- 131.

\bibitem[Brouwer and Haemers, 2011]{brouwer2011spectra}
Brouwer, A.~E. and Haemers, W.~H. (2011).
\newblock {\em Spectra of graphs}.
\newblock Springer Science \& Business Media.

\bibitem[Brynjarsd{\'o}ttir and Gelfand, 2014]{brynjarsdottir2014covariate}
Brynjarsd{\'o}ttir, J. and Gelfand, A.~E. (2014).
\newblock On covariate importance for regression models with multivariate response.
\newblock {\em Journal of Agricultural, Biological, and Environmental Statistics}, 19(4):479--500.

\bibitem[{Center for Space Research, University of Texas at Austin}, 2025]{GraceMascons}
{Center for Space Research, University of Texas at Austin} (2025).
\newblock {CSR GRACE/GRACE-FO RL06.3 Mascon Solutions (RL0603)}.
\newblock \url{https://www2.csr.utexas.edu/grace/RL06_mascons.html}.

\bibitem[Cressie et~al., 2022]{cressie2022basis}
Cressie, N., Sainsbury-Dale, M., and Zammit-Mangion, A. (2022).
\newblock Basis-function models in spatial statistics.
\newblock {\em Annual Review of Statistics and Its Application}, 9(1):373--400.

\bibitem[Dark and Bram, 2007]{MAUPGeography}
Dark, S.~J. and Bram, D. (2007).
\newblock The modifiable areal unit problem (maup) in physical geography.
\newblock {\em Progress in Physical Geography: Earth and Environment}, 31(5):471--479.

\bibitem[Deines et~al., 2019]{deines2019quantifying}
Deines, J.~M., Kendall, A.~D., Butler, J.~J., and Hyndman, D.~W. (2019).
\newblock Quantifying irrigation adaptation strategies in response to stakeholder-driven groundwater management in the us high plains aquifer.
\newblock {\em Environmental Research Letters}, 14(4):044014.

\bibitem[Everett, 2013]{everett2013introduction}
Everett, B. (2013).
\newblock {\em An introduction to latent variable models}.
\newblock Springer Science \& Business Media.

\bibitem[Fortin et~al., 2012]{fortin2012spatial}
Fortin, M.-J., James, P.~M., MacKenzie, A., Melles, S.~J., and Rayfield, B. (2012).
\newblock Spatial statistics, spatial regression, and graph theory in ecology.
\newblock {\em Spatial Statistics}, 1:100--109.

\bibitem[Griffith, 2003]{griffith2003spatial}
Griffith, D.~A. (2003).
\newblock Spatial filtering.
\newblock In {\em Spatial autocorrelation and spatial filtering: Gaining understanding through theory and scientific visualization}, pages 91--130. Springer.

\bibitem[Haacker et~al., 2016]{Haacker2016-to}
Haacker, E. M.~K., Kendall, A.~D., and Hyndman, D.~W. (2016).
\newblock Water level declines in the high plains aquifer: Predevelopment to resource senescence.
\newblock {\em Ground Water}, 54(2):231--242.

\bibitem[Hancock, 2024]{hancock2024texas}
Hancock, K. (2024).
\newblock Texas comptroller high plain region 2024 report.
\newblock \url{https://comptroller.texas.gov/economy/economic-data/regions/2024/high-plains.php}.

\bibitem[Haqiqi et~al., 2023]{haqiqi2023global}
Haqiqi, I., Bowling, L., Jame, S., Baldos, U., Liu, J., and Hertel, T. (2023).
\newblock Global drivers of local water stresses and global responses to local water policies in the united states.
\newblock {\em Environmental Research Letters}, 18(6):065007.

\bibitem[Haran, 2011]{haran2011gaussian}
Haran, M. (2011).
\newblock Gaussian random field models for spatial data.
\newblock {\em Handbook of Markov Chain Monte Carlo}, pages 449--478.

\bibitem[Haran and Lee, 2024]{haran2024latent}
Haran, M. and Lee, B. (2024).
\newblock Latent gaussian models and computation for large spatial data.
\newblock {\em Handbook of Markov Chain Monte Carlo, Handbooks of Modern Statistical Methods. Chapman and Hall/CRC}.

\bibitem[Hazra et~al., 2023]{Hazra2023}
Hazra, A., Huser, R., and J{\'o}hannesson, {\'A}.~V. (2023).
\newblock {\em Bayesian Latent Gaussian Models for High-Dimensional Spatial Extremes}, pages 219--251.
\newblock Springer International Publishing, Cham.

\bibitem[Hector and Martin, 2024]{hector2024efficient}
Hector, E.~C. and Martin, R. (2024).
\newblock Turning the information-sharing dial: efficient inference from different data sources.

\bibitem[Hrafnkelsson and Bakka, 2023]{Hrafnkelsson2023}
Hrafnkelsson, B. and Bakka, H. (2023).
\newblock {\em Bayesian Latent Gaussian Models}, pages 1--80.
\newblock Springer International Publishing, Cham.

\bibitem[Hughes, 2017]{hughes2017spatial}
Hughes, J. (2017).
\newblock Spatial regression and the bayesian filter.
\newblock {\em arXiv preprint arXiv:1706.04651}.

\bibitem[Hughes and Haran, 2013]{hughes2013dimension}
Hughes, J. and Haran, M. (2013).
\newblock Dimension reduction and alleviation of confounding for spatial generalized linear mixed models.
\newblock {\em Journal of the Royal Statistical Society Series B: Statistical Methodology}, 75(1):139--159.

\bibitem[Isaaks and Srivastava, 1989]{Isaaks1989AppliedGeostatistics}
Isaaks, E.~H. and Srivastava, R.~M. (1989).
\newblock {\em An Introduction to Applied Geostatistics}.
\newblock Oxford University Press, New York, NY.

\bibitem[Jacob, 2015]{jacob2012ebsco}
Jacob, L. (2015).
\newblock 2012-2015 north american drought.
\newblock \url{https://www.ebsco.com/research-starters/history/2012-2015-north-american-drought}.

\bibitem[Jelinski and Wu, 1996]{Jelinski1996}
Jelinski, D.~E. and Wu, J. (1996).
\newblock The modifiable areal unit problem and implications for landscape ecology.
\newblock {\em Landscape Ecology}, 11(3):129--140.

\bibitem[K{\"u}nsch, 1979]{kunsch1979gaussian}
K{\"u}nsch, H. (1979).
\newblock Gaussian markov random fields.
\newblock {\em J. Fac. Sci. Univ. Tokyo Sect. IA Math}, 26(1):53--73.

\bibitem[Landerer et~al., 2020]{Landerer2020}
Landerer, F.~W., Flechtner, F.~M., Save, H., Webb, F.~H., Bandikova, T., Bertiger, W.~I., Bettadpur, S.~V., Byun, S.~H., Dahle, C., Dobslaw, H., Fahnestock, E., Harvey, N., Kang, Z., Kruizinga, G. L.~H., Loomis, B.~D., McCullough, C., Murböck, M., Nagel, P., Paik, M., Pie, N., Poole, S., Strekalov, D., Tamisiea, M.~E., Wang, F., Watkins, M.~M., Wen, H.-Y., Wiese, D.~N., and Yuan, D.-N. (2020).
\newblock Extending the global mass change data record: Grace follow-on instrument and science data performance.
\newblock {\em Geophysical Research Letters}, 47(12):e2020GL088306.
\newblock e2020GL088306 2020GL088306.

\bibitem[Li et~al., 2023]{li2023combined}
Li, C.-H., Mao, J.-J., Wu, Y.-J., Zhang, B., Zhuang, X., Qin, G., and Liu, H.-M. (2023).
\newblock {Combined impacts of environmental and socioeconomic covariates on HFMD risk in China: A spatiotemporal heterogeneous perspective}.
\newblock {\em PLOS Neglected Tropical Diseases}, 17(5):e0011286.

\bibitem[Liu et~al., 2022]{Liu2022SpecYield}
Liu, G., Wilson, B.~B., Bohling, G.~C., Whittemore, D.~O., and Butler~Jr, J.~J. (2022).
\newblock Estimation of specific yield for regional groundwater models: Pitfalls, ramifications, and a promising path forward.
\newblock {\em Water Resources Research}, 58(1):e2021WR030761.

\bibitem[Lohr and Raghunathan, 2017]{LohrRaghunathan}
Lohr, S.~L. and Raghunathan, T.~E. (2017).
\newblock {Combining Survey Data with Other Data Sources}.
\newblock {\em Statistical Science}, 32(2):293 -- 312.

\bibitem[Manski, 2001]{manski2001designing}
Manski, C.~F. (2001).
\newblock Designing programs for heterogeneous populations: The value of covariate information.
\newblock {\em American Economic Review}, 91(2):103--106.

\bibitem[Mehrnegar et~al., 2023]{mehrnegar2023making}
Mehrnegar, N., Schumacher, M., Jagdhuber, T., and Forootan, E. (2023).
\newblock Making the best use of grace, grace-fo and smap data through a constrained bayesian data-model integration.
\newblock {\em Water Resources Research}, 59(9):e2023WR034544.

\bibitem[Mertens and Lambin, 1997]{mertens1997spatial}
Mertens, B. and Lambin, E.~F. (1997).
\newblock {Spatial modelling of deforestation in southern Cameroon: spatial disaggregation of diverse deforestation processes}.
\newblock {\em Applied Geography}, 17(2):143--162.

\bibitem[Monteiro et~al., 2019]{ijgi8080327}
Monteiro, J., Martins, B., Murrieta-Flores, P., and Pires, J.~M. (2019).
\newblock Spatial disaggregation of historical census data leveraging multiple sources of ancillary information.
\newblock {\em ISPRS International Journal of Geo-Information}, 8(8).

\bibitem[Moran, 1950]{moran1950notes}
Moran, P.~A. (1950).
\newblock Notes on continuous stochastic phenomena.
\newblock {\em Biometrika}, 37(1/2):17--23.

\bibitem[Muhling et~al., 2018]{muhling2018potential}
Muhling, B.~A., Gait{\'a}n, C.~F., Stock, C.~A., Saba, V.~S., Tommasi, D., and Dixon, K.~W. (2018).
\newblock Potential salinity and temperature futures for the chesapeake bay using a statistical downscaling spatial disaggregation framework.
\newblock {\em Estuaries and Coasts}, 41:349--372.

\bibitem[Paige et~al., 2022]{paige2022spatial}
Paige, J., Fuglstad, G.-A., Riebler, A., and Wakefield, J. (2022).
\newblock Spatial aggregation with respect to a population distribution: Impact on inference.
\newblock {\em Spatial Statistics}, 52:100714.

\bibitem[Pakrashi et~al., 2025]{Pakrashi2025}
Pakrashi, A., Hazra, A., Raveendran, S.~M., and Balakrishnan, K. (2025).
\newblock Approximate {B}ayesian inference for high-resolution spatial disaggregation using alternative data sources.
\newblock {\em Journal of Agricultural, Biological and Environmental Statistics}, 30(2):576--599.

\bibitem[Pervez and Brown, 2010]{pervez2010irrigation}
Pervez, M.~S. and Brown, J.~F. (2010).
\newblock Mapping irrigated lands at 250-m scale by merging modis data and national agricultural statistics.
\newblock {\em Remote Sensing}, 2(10):2388--2412.

\bibitem[Pollet et~al., 2015]{pollet2015taking}
Pollet, T.~V., Stulp, G., Henzi, S.~P., and Barrett, L. (2015).
\newblock Taking the aggravation out of data aggregation: A conceptual guide to dealing with statistical issues related to the pooling of individual-level observational data.
\newblock {\em American journal of primatology}, 77(7):727--740.

\bibitem[Ponciano et~al., 2009]{ponciano2009hierarchical}
Ponciano, J.~M., Taper, M.~L., Dennis, B., and Lele, S.~R. (2009).
\newblock Hierarchical models in ecology: confidence intervals, hypothesis testing, and model selection using data cloning.
\newblock {\em Ecology}, 90(2):356--362.

\bibitem[Qi, 2010]{Qi2010HighPlains}
Qi, S. (2010).
\newblock Digital map of the aquifer boundary for the high plains aquifer in parts of colorado, kansas, nebraska, new mexico, oklahoma, south dakota, texas, and wyoming.

\bibitem[Rhodes et~al., 2023]{RHODES202383}
Rhodes, E.~C., Perotto-Baldivieso, H.~L., Tanner, E.~P., Angerer, J.~P., and Fox, W.~E. (2023).
\newblock The declining ogallala aquifer and the future role of rangeland science on the north american high plains.
\newblock {\em Rangeland Ecology \& Management}, 87:83--96.

\bibitem[Rippey, 2015]{RIPPEY201557}
Rippey, B.~R. (2015).
\newblock The u.s. drought of 2012.
\newblock {\em Weather and Climate Extremes}, 10:57--64.
\newblock USDA Research and Programs on Extreme Events.

\bibitem[Roquette et~al., 2018]{roquette2018relevance}
Roquette, R., Nunes, B., and Painho, M. (2018).
\newblock {The relevance of spatial aggregation level and of applied methods in the analysis of geographical distribution of cancer mortality in mainland Portugal (2009--2013)}.
\newblock {\em Population health metrics}, 16:1--12.

\bibitem[Rudstrom et~al., 2002]{rudstrom2002data}
Rudstrom, M., Popp, M., Manning, P., and Gbur, E. (2002).
\newblock Data aggregation issues for crop yield risk analysis.
\newblock {\em Canadian Journal of Agricultural Economics/Revue canadienne d'agroeconomie}, 50(2):185--200.

\bibitem[Rue and Held, 2005]{rue2005gaussian}
Rue, H. and Held, L. (2005).
\newblock {\em Gaussian Markov random fields: theory and applications}.
\newblock Chapman and Hall/CRC.

\bibitem[Scanlon et~al., 2012]{scanlon2012groundwater}
Scanlon, B.~R., Faunt, C.~C., Longuevergne, L., Reedy, R.~C., Alley, W.~M., McGuire, V.~L., and McMahon, P.~B. (2012).
\newblock Groundwater depletion and sustainability of irrigation in the us high plains and central valley.
\newblock {\em Proceedings of the national academy of sciences}, 109(24):9320--9325.

\bibitem[Schmid and Brown, 2000]{schmid2000bayesian}
Schmid, C.~H. and Brown, E.~N. (2000).
\newblock Bayesian hierarchical models.
\newblock {\em Methods in enzymology}, 321:305--330.

\bibitem[Segond et~al., 2007]{segond2007simulation}
Segond, M.-L., Neokleous, N., Makropoulos, C., Onof, C., and Maksimovic, C. (2007).
\newblock Simulation and spatio-temporal disaggregation of multi-site rainfall data for urban drainage applications.
\newblock {\em Hydrological sciences journal}, 52(5):917--935.

\bibitem[Shrestha et~al., 2019]{shrestha2019moderate}
Shrestha, D., Howard, D., and Benedict, T.~D. (2019).
\newblock Moderate resolution imaging spectroradiometer (modis) irrigated agriculture datasets for the conterminous united states (mirad-us).
\newblock {\em US Geological Survey (USGS) Data Release}, page 920.

\bibitem[Sophocleous, 2005]{sophocleous2005groundwater}
Sophocleous, M. (2005).
\newblock Groundwater recharge and sustainability in the high plains aquifer in kansas, usa.
\newblock {\em Hydrogeology Journal}, 13(2):351--365.

\bibitem[Stanton et~al., 2010]{stanton2010simulation}
Stanton, J.~S., Peterson, S.~M., and Fienen, M.~N. (2010).
\newblock Simulation of groundwater flow and effects of groundwater irrigation on stream base flow in the elkhorn and loup river basins, nebraska, 1895--2055---phase two.
\newblock Scientific Investigations Report 2010-5149, U.S. Geological Survey, Reston, VA.

\bibitem[Steward and Allen, 2016]{STEWARD201636}
Steward, D.~R. and Allen, A.~J. (2016).
\newblock Peak groundwater depletion in the high plains aquifer, projections from 1930 to 2110.
\newblock {\em Agricultural Water Management}, 170:36--48.
\newblock Special Issue: Water Management Strategies in Irrigated Areas Overseen by: Dr. Brent Clothier.

\bibitem[Steward et~al., 2013]{steward2013tapping}
Steward, D.~R., Bruss, P.~J., Yang, X., Staggenborg, S.~A., Welch, S.~M., and Apley, M.~D. (2013).
\newblock Tapping unsustainable groundwater stores for agricultural production in the high plains aquifer of kansas, projections to 2110.
\newblock {\em Proceedings of the National Academy of Sciences}, 110(37):E3477--E3486.

\bibitem[Tapley et~al., 2019]{Tapley2019-im}
Tapley, B.~D., Watkins, M.~M., Flechtner, F., Reigber, C., Bettadpur, S., Rodell, M., Sasgen, I., Famiglietti, J.~S., Landerer, F.~W., Chambers, D.~P., Reager, J.~T., Gardner, A.~S., Save, H., Ivins, E.~R., Swenson, S.~C., Boening, C., Dahle, C., Wiese, D.~N., Dobslaw, H., Tamisiea, M.~E., and Velicogna, I. (2019).
\newblock Contributions of {GRACE} to understanding climate change.
\newblock {\em Nat. Clim. Chang.}, 5(5):358--369.

\bibitem[Utazi et~al., 2019]{utazi2019spatial}
Utazi, C., Thorley, J., Alegana, V., Ferrari, M., Nilsen, K., Takahashi, S., Metcalf, C. J.~E., Lessler, J., and Tatem, A. (2019).
\newblock A spatial regression model for the disaggregation of areal unit based data to high-resolution grids with application to vaccination coverage mapping.
\newblock {\em Statistical Methods in Medical Research}, 28(10-11):3226--3241.

\bibitem[van Beurden and Douven, 1999]{van1999aggregation}
van Beurden, A.~U. and Douven, W.~J. (1999).
\newblock Aggregation issues of spatial information in environmental research.
\newblock {\em International Journal of Geographical Information Science}, 13(5):513--527.

\bibitem[Wahr et~al., 1998]{Wahr1998-jn}
Wahr, J., Molenaar, M., and Bryan, F. (1998).
\newblock Time variability of the earth's gravity field: Hydrological and oceanic effects and their possible detection using {GRACE}.
\newblock {\em J. Geophys. Res.}, 103(B12):30205--30229.

\bibitem[Wainwright et~al., 2016]{wainwright2016hierarchical}
Wainwright, H.~M., Flores~Orozco, A., B{\"u}cker, M., Dafflon, B., Chen, J., Hubbard, S.~S., and Williams, K.~H. (2016).
\newblock {Hierarchical Bayesian method for mapping biogeochemical hot spots using induced polarization imaging}.
\newblock {\em Water Resources Research}, 52(1):533--551.

\bibitem[Wikle, 2003]{wikle2003hierarchical}
Wikle, C.~K. (2003).
\newblock Hierarchical bayesian models for predicting the spread of ecological processes.
\newblock {\em Ecology}, 84(6):1382--1394.

\bibitem[Wikle, 2010]{wikle2010low}
Wikle, C.~K. (2010).
\newblock Low-rank representations for spatial processes.
\newblock {\em Handbook of spatial statistics}, 107:118.

\bibitem[Wikle et~al., 1998]{wikle1998hierarchical}
Wikle, C.~K., Berliner, L.~M., and Cressie, N. (1998).
\newblock Hierarchical bayesian space-time models.
\newblock {\em Environmental and ecological statistics}, 5(2):117--154.

\bibitem[Williams and Rasmussen, 2006]{williams2006gaussian}
Williams, C.~K. and Rasmussen, C.~E. (2006).
\newblock {\em Gaussian processes for machine learning}, volume~2.
\newblock MIT press Cambridge, MA.

\bibitem[Wireman, 2015]{Wireman2015-ey}
Wireman, M. (2015).
\newblock Development and implementation of a national groundwater monitoring network.
\newblock {\em Ground Water Monit. Remediat.}, 35(2):21--22.

\bibitem[Wunsch and Schreiber, 2024]{WunschSchreiber2024}
Wunsch, D.~R. and Schreiber, R.~P. (2024).
\newblock Calling all groundwater professionals: Support the national groundwater monitoring network.
\newblock {\em Groundwater}, 62(3):328--329.

\bibitem[Yang and Ding, 2020]{Yang02072020}
Yang, S. and Ding, P. (2020).
\newblock Combining multiple observational data sources to estimate causal effects.
\newblock {\em Journal of the American Statistical Association}, 115(531):1540--1554.
\newblock PMID: 33088006.

\bibitem[Yang et~al., 2022]{yang2022classification}
Yang, Y., Gao, H., Berry, C., Carrick, D., Radjenovic, A., and Husmeier, D. (2022).
\newblock {Classification of myocardial blood flow based on dynamic contrast-enhanced magnetic resonance imaging using hierarchical Bayesian models}.
\newblock {\em Journal of the Royal Statistical Society Series C: Applied Statistics}, 71(5):1085--1115.

\bibitem[You et~al., 2009]{you2009generating}
You, L., Wood, S., and Wood-Sichra, U. (2009).
\newblock {Generating plausible crop distribution maps for Sub-Saharan Africa using a spatially disaggregated data fusion and optimization approach}.
\newblock {\em Agricultural Systems}, 99(2-3):126--140.

\end{thebibliography}
